\documentclass[aip,jcp,preprint,groupedaddress]{revtex4-2}%
\usepackage{graphicx}
\usepackage{tabularx}
\usepackage{dcolumn}
\usepackage{longtable}
\usepackage{tensor}
\usepackage{xcolor}
\usepackage{placeins}
\usepackage{bm}
\usepackage{amsmath}
\usepackage{amsfonts}
\usepackage{amssymb}
\usepackage{float}%
\providecommand{\U}[1]{\protect\rule{.1in}{.1in}}
\providecommand{\U}[1]{\protect\rule{.1in}{.1in}}
\providecommand{\U}[1]{\protect\rule{.1in}{.1in}}
\begin{document}
\title{Which form of the molecular Hamiltonian is the most suitable for simulating
the nonadiabatic quantum dynamics at a conical intersection?}
\author{Seonghoon Choi}
\email{seonghoon.choi@epfl.ch}
\author{Ji\v{r}\'{\i} Van\'{\i}\v{c}ek}
\email{jiri.vanicek@epfl.ch}
\affiliation{Laboratory of Theoretical Physical Chemistry, Institut des Sciences et
Ing\'enierie Chimiques, Ecole Polytechnique F\'ed\'erale de Lausanne (EPFL),
CH-1015, Lausanne, Switzerland}
\date{\today}

\begin{abstract}
Choosing an appropriate representation of the molecular Hamiltonian is one of
the challenges faced by simulations of the nonadiabatic quantum dynamics
around a conical intersection. The adiabatic, exact quasidiabatic, and
strictly diabatic representations are exact and unitary transforms of each
other, whereas the approximate quasidiabatic Hamiltonian ignores the residual
nonadiabatic couplings in the exact quasidiabatic Hamiltonian. A rigorous
numerical comparison of the four different representations is difficult
because of the exceptional nature of systems where the four representations
can be defined exactly and the necessity of an exceedingly accurate numerical
algorithm that avoids mixing numerical errors with errors due to the different
forms of the Hamiltonian. Using the quadratic Jahn-Teller model and high-order
geometric integrators, we are able to perform this comparison and find that
only the rarely employed exact quasidiabatic Hamiltonian yields nearly
identical results to the benchmark results of the strictly diabatic
Hamiltonian, which is not available in general. In this Jahn-Teller model and
with the same Fourier grid, the commonly employed approximate quasidiabatic
Hamiltonian led to inaccurate wavepacket dynamics, while the Hamiltonian in
the adiabatic basis was the least accurate, due to the singular nonadiabatic
couplings at the conical intersection.

\end{abstract}
\maketitle

\graphicspath{{./figures/}{C:/Users/Jiri/Dropbox/Papers/Chemistry_papers/2020/Hamiltonian_representation_CI/figures/}{"d:/Group Vanicek/Desktop/Choi/Residual_coupling/figures/"}}

\section{\label{sec:introduction}Introduction}
Many physical and chemical phenomena proceed via conical
intersections---nuclear geometries where adiabatic potential energy surfaces
of two or more electronic states intersect.\cite{Teller:1937, Forster:1970,
Herzberg_LonguetHiggins:1963, Zimmerman:1966, Koppel_Cederbaum:1984} The
conical intersections, which are much more ubiquitous\cite{Yarkony:1990,
Atchity_Ruedenberg:1991, Bernardi_Robb:1990, book_Klessinger_Michl:1995} than
previously believed, are responsible for the failure of the celebrated
Born--Oppenheimer approximation that treats the electronic and nuclear motions
in molecules separately. To correctly describe processes involving conical
intersections, methods going beyond\cite{book_Baer:2006,Domcke_Yarkony:2012,
book_Nakamura:2012, book_Takatsuka:2015, Bircher_Rothlisberger:2017,
Shin_Metiu:1995, Albert_Engel:2016, Abedi_Gross:2010, Cederbaum:2008} the
Born--Oppenheimer approximation must be used. Often, it is necessary to take
into account multiple strongly coupled electronic
states\cite{Worth_Cederbaum:2004, book_Baer:2006, Cederbaum:2004} and, in the
adiabatic basis, one must consider the geometric phase effect---the sign
change of adiabatic electronic states along a closed path encircling a conical
intersection.\cite{Longuet-Higgins_Sack:1958, Mead_Truhlar:1979, Berry:1984,
Mead:1992, Kendrick:2000, Juanes-Marcos_Althorpe:2005, Malbon_Yarkony:2016,
Xie_Yarkony:2019, Xie_Guo:2017, Ryabinkin_Izmailov:2017,
Joubert-Dorial_Izmaylov:2013, Schon_Koppel:1995}

Simulating the nonadiabatic quantum dynamics at a conical intersection using
the adiabatic Hamiltonian, obtained from the electronic structure
calculations, is problematic due to both the geometric phase
effect\cite{Longuet-Higgins_Sack:1958, Mead:1992, Malbon_Yarkony:2016,
Xie_Yarkony:2019, Xie_Guo:2017, Ryabinkin_Izmailov:2017,
Joubert-Dorial_Izmaylov:2013, Schon_Koppel:1995} and singular nonadiabatic
couplings.\cite{Cederbaum:2004} These are rectified by a unitary
transformation into the equivalent exact quasidiabatic Hamiltonian. Unlike the
adiabatic electronic states, which are only coupled through the nonadiabatic
couplings, the quasidiabatic states have both the residual, presumably small,
nonadiabatic couplings and the diabatic couplings---the off-diagonal elements
of the potential energy matrix. However, the exact quasidiabatic Hamiltonian
is rarely used. Instead, the approximate quasidiabatic Hamiltonian, which
ignores the residual nonadiabatic couplings, is almost always used due to its
simplicity. The separable form of the approximate quasidiabatic Hamiltonian
allows using a wider range of time propagation
schemes,\cite{Leforestier_Kosloff:1991, Tal-Ezer_Kosloff:1984,
Park_Light:1986, book_Hairer_Wanner:2006} including the well-known
split-operator algorithm,\cite{Feit_Steiger:1982, book_Lubich:2008,
book_Tannor:2007} but ignoring the residual nonadiabatic couplings
decreases the accuracy.\cite{Choi_Vanicek:2020} The strictly diabatic
Hamiltonian, with only diabatic couplings and no nonadiabatic couplings, would
be the most suitable for the nonadiabatic quantum dynamics simulation.
However, in typical systems, the strictly diabatic states only exist in
general when an infinite number of electronic states are
considered.\cite{Mead_Truhlar:1982, Pacher_Koppel:1989}

Advantages and disadvantages of various Hamiltonians have been explored by
numerous comparisons of the nonadiabatic quantum dynamics simulated with
different Hamiltonians: To name a few, there exist comparisons between the
strictly diabatic and approximate quasidiabatic
Hamiltonians,\cite{Koppel_Mahapatra:2001, Pacher_Koppel:1989,
Pacher_Koppel:1988, Gadea_Pelissier:1990, Thiel_Koppel:1999} between the
adiabatic and approximate quasidiabatic Hamiltonians,\cite{Viel_Domcke:2004,
Ben-Nun_Martinez:2000, Ibele_Curchod:2020} between the adiabatic and exact
quasidiabatic Hamiltonians,\cite{Mandal_Huo:2018, Zhou_Huo:2019} and between
the adiabatic and strictly diabatic Hamiltonians.\cite{Guo_Yarkony:2016,
Xie_Guo:2016, Joubert-Dorial_Izmaylov:2013} To the best of our knowledge, no
study has compared all four different Hamiltonians---the adiabatic, exact
quasidiabatic, approximate quasidiabatic, and strictly diabatic
Hamiltonians---on a single system. A rigorous comparison is challenging
because one must have both an appropriate system, where the different
Hamiltonians are defined exactly, and a highly-accurate numerical integrator,
which allows separating numerical errors from the errors due to using
different forms of the Hamiltonian.

The two-dimensional, two-state quadratic $E \otimes e$ Jahn--Teller
model\cite{book_Bersuker_Polinger:2012, Bersuker:2001, Thiel_Koppel:1999} is
perfect for this comparison because analytical expressions exist for potential
energy surfaces and nonadiabatic couplings in both the adiabatic and
quasidiabatic representations and because this model has---exceptionally---a
strictly diabatic Hamiltonian.

As for numerical integrators, the split-operator
algorithms\cite{Feit_Steiger:1982, book_Lubich:2008, book_Tannor:2007} are
applicable to both the approximate quasidiabatic and strictly diabatic
Hamiltonians because these Hamiltonians are separable, i.e., they can be
expressed as sums of terms depending purely on either the position or momentum
operator. In contrast, neither the adiabatic nor exact quasidiabatic
Hamiltonian is separable due to the nonvanishing nonadiabatic couplings.
Although the split-operator algorithms cannot be used, the wavepacket can be
propagated with the implicit midpoint method.\cite{book_Hairer_Wanner:2006,
book_Leimkuhler_Reich:2004} Both the split-operator and implicit midpoint
methods preserve most geometric properties of the exact solution (such as norm
conservation, stability, and time reversibility) and, in addition, can be
symmetrically composed\cite{Suzuki:1990, Yoshida:1990, Kahan_Li:1997,
book_Hairer_Wanner:2006} to obtain integrators of arbitrary even order in the
time step.\cite{Choi_Vanicek:2019, Roulet_Vanicek:2019} By taking advantage of
the suitable model and high-order geometric integrators, we numerically
compared the wavepacket and observables obtained from simulations with the
different Hamiltonians: the adiabatic, exact quasidiabatic, approximate
quasidiabatic, and strictly diabatic Hamiltonians.

\section{\label{sec:theory}Theory}

The molecular Hamiltonian can be partitioned as $\mathcal{H}=\mathcal{T}%
_{\mathrm{N}}+\mathcal{H}_{\mathrm{e}}(Q)$, where $\mathcal{T}_{\mathrm{N}}$
is the nuclear kinetic energy operator and $\mathcal{H}_{\mathrm{e}}(Q)$
denotes the electronic Hamiltonian, which depends parametrically on the
$D$-dimensional vector $Q$ of nuclear coordinates. Using the adiabatic
electronic states $|n(Q)\rangle$, obtained by solving the time-independent
Schr\"{o}dinger equation
\begin{equation}
\mathcal{H}_{\mathrm{e}}(Q)|n(Q)\rangle=V_{n}(Q)|n(Q)\rangle
,\label{eq:electronic_tise}%
\end{equation}
one can establish an approximate ansatz 
\begin{align}
|\Psi(Q,t)\rangle &  =\sum_{n=1}^{S}\psi_{n}(Q,t)|n(Q)\rangle\nonumber\\
&  =\sum_{n=1}^{S}[\psi_{n}(Q,t)e^{-iA_{n}(Q)}][e^{iA_{n}(Q)}|n(Q)\rangle
]\label{eq:tdse_approx_ansatz}%
\end{align}
 for the solution of the time-dependent Schr\"{o}dinger equation
\begin{equation}
i\hbar\frac{\partial}{\partial t}|\Psi(Q,t)\rangle=\mathcal{H}|\Psi
(Q,t)\rangle;\label{eq:tdse}%
\end{equation}
 here, $\psi_{n}(Q,t)$, $V_{n}(Q)$, and $e^{iA_{n}(Q)}$ are the
time-dependent nuclear wavefunction, potential energy surface, and
coordinate-dependent phase factor associated with the $n$th adiabatic
electronic state. The Born--Huang expansion\cite{book_Born_Huang:1954} in
Eq.~(\ref{eq:tdse_approx_ansatz}) is not exact unless the sum includes an
infinite number of terms but can be very accurate if a finite number $S$ of
electronic states are chosen wisely.\cite{Ballhausen_Hansen:1972,
Yarkony:1996, book_Baer:2006}

One is free to choose an overall phase $A_{n}(Q)$ in
Eq.~(\ref{eq:tdse_approx_ansatz}) because if $|n(Q)\rangle$ is a normalized
solution of Eq.~(\ref{eq:electronic_tise}), then so is $e^{iA_{n}%
(Q)}|n(Q)\rangle$. Unless $A_{n}(Q)$ is carefully chosen, however, the
adiabatic states and, therefore, also the wavepackets undergo a sign change
along a closed path encircling a conical
intersection.\cite{Longuet-Higgins_Sack:1958, Herzberg_LonguetHiggins:1963,
Mead_Truhlar:1979, Berry:1984, Mead:1992, Kendrick:2000,
Juanes-Marcos_Althorpe:2005, Schon_Koppel:1995, Malbon_Yarkony:2016,
Xie_Yarkony:2019, Xie_Guo:2017, Ryabinkin_Izmailov:2017,
Joubert-Dorial_Izmaylov:2013, Xie_Guo:2017} In Sec. S4 of the supplementary
material, we show that neglecting this double-valuedness of the wavepackets is
detrimental to accuracy. Instead, in what follows, we set phases $A_{n}{(Q)}$
appropriately (see Sec. S1 of the supplementary material) to ensure the
single-valuedness of both the adiabatic states and the wavepackets. In other
words, we include the \textquotedblleft geometric phase\textquotedblright\ in
the adiabatic states in order to obtain the best possible results in the
adiabatic representation. From now on, we absorb the overall phase factors
$e^{-iA_{n}(Q)}$ and $e^{iA_{n}(Q)}$ into the nuclear wavefunctions $\psi
_{n}(Q,t)$ and adiabatic states $|n(Q)\rangle$.

The time-dependent Schr\"{o}dinger equation in the adiabatic representation,
\begin{equation}
i\hbar\frac{d}{dt}\bm{\psi}(t)=\hat{\mathbf{H}}_{\mathrm{ad}}\bm{\psi}(t),
\label{eq:tdse_adiab}%
\end{equation}
is obtained by substituting ansatz~(\ref{eq:tdse_approx_ansatz}) into
Eq.~(\ref{eq:tdse}) and projecting onto electronic states $\langle m(Q)|$ for
$m\in\{1,\dots,S\}$. Note that we have introduced the
representation-independent matrix notation: \textbf{bold} font indicates
either an $S\times S$ matrix, i.e., an electronic operator, or $S$-dimensional
vector, and the hat $\hat{}$ indicates a nuclear operator. In particular,
$(\hat{\mathbf{H}}_{\mathrm{ad}})_{mn}=\langle m|\mathcal{H}|n\rangle$ is the
adiabatic Hamiltonian, and $\bm{\psi}(t)$ denotes the molecular wavepacket in
the adiabatic representation with components $\psi_{n}(t)$; henceforth,
$m,n\in\{1,\dots,S\}$ unless otherwise stated. The formal solution of
Eq.~(\ref{eq:tdse_adiab}) for a given initial condition $\bm{\psi}(0)$ is
\begin{equation}
\bm{\psi}_{\mathrm{ad}}(t):=\hat{\mathbf{U}}_{\mathrm{ad}}(t)\bm{\psi}(0),
\label{eq:tdse_solution_ad}%
\end{equation}
where $\hat{\mathbf{U}}_{\mathrm{ad}}(t)$ denotes the exact evolution operator
$\hat{\mathbf{U}}_{i}(t):=\exp(-i\hat{\mathbf{H}}_{i}t/\hbar)$ with
$i=\mathrm{ad}$. The adiabatic Hamiltonian $\hat{\mathbf{H}}_{\mathrm{ad}}$ is
often expressed as
\begin{equation}
\hat{\mathbf{H}}_{\mathrm{ad}}=\frac{1}{2M}[\hat{P}^{2}\mathbf{1}%
-2i\hbar\mathbf{F}_{\mathrm{ad}}(\hat{Q})\cdot\hat{P}-\hbar^{2}\mathbf{G}%
_{\mathrm{ad}}(\hat{Q})]+\mathbf{V}_{\mathrm{ad}}(\hat{Q}),
\label{eq:mol_H_adiab}%
\end{equation}
where we have used the matrix notation for the diagonal adiabatic potential
energy matrix $[\mathbf{V}_{\mathrm{ad}}(Q)]_{mn}:=V_{n}(Q)\delta_{mn}$,
nonadiabatic vector couplings $[\mathbf{F}_{\mathrm{ad}}(Q)]_{mn}:=\langle
m(Q)|\nabla n(Q)\rangle$, and nonadiabatic scalar couplings $[\mathbf{G}%
_{\mathrm{ad}}(Q)]_{mn}:=\langle m(Q)|\nabla^{2}n(Q)\rangle$. The
$D$-dimensional vector $P$ is the nuclear momentum conjugate to $Q$ and the
dot $\cdot$ denotes a dot product in the $D$-dimensional nuclear vector space;
we use the mass-scaled coordinates for simplicity.

Expressing the nonadiabatic vector couplings as
\begin{equation}
\lbrack\mathbf{F}_{\mathrm{ad}}(Q)]_{mn}=\frac{\langle m(Q)|\nabla
\mathcal{H}_{\mathrm{e}}(Q)|n(Q)\rangle}{V_{n}(Q)-V_{m}(Q)},\quad m\neq
n,\label{eq:F_ad_hellman_feynmann}%
\end{equation}
shows that they are singular at a conical intersection,\cite{Koppel:2004a}
which is a nuclear geometry $Q_{0}$ where $V_{m}(Q_{0})=V_{n}(Q_{0})$ for
$m\neq n$.\cite{Domcke_Yarkony:2012, Cederbaum:2004, Yarkony:2004} This
singularity causes problems for the nonadiabatic dynamics simulations, and
especially for the grid-based methods because an infinitely dense grid would
be required to describe the singularity. The singularity, however, can be
removed by a coordinate-dependent unitary transformation of
Hamiltonian~(\ref{eq:mol_H_adiab}) into its quasidiabatic representation,
\begin{align}
&  \hat{\mathbf{H}}_{\text{qd-exact}}=\mathbf{S}(\hat{Q})\hat{\mathbf{H}%
}_{\mathrm{ad}}\mathbf{S}(\hat{Q})^{\dagger}\nonumber\\
&  =\frac{1}{2M}[\hat{P}^{2}\mathbf{1}-2i\hbar\mathbf{F}_{\mathrm{qd}}(\hat
{Q})\cdot\hat{P}-\hbar^{2}\mathbf{G}_{\mathrm{qd}}(\hat{Q})]+\mathbf{V}%
_{\mathrm{qd}}(\hat{Q}),\label{eq:mol_H_qd}%
\end{align}
where $\mathbf{F}_{\mathrm{qd}}(Q)$ and $\mathbf{G}_{\mathrm{qd}}(Q)$ denote
the residual nonadiabatic vector and scalar couplings, respectively. Different
transformation matrices $\mathbf{S}(Q)$ are obtained by different
quasidiabatization schemes,\cite{Koppel:2004a} which include the
block-diagonalization of the reference Hamiltonian
matrix,\cite{Pacher_Koppel:1988, Pacher_Cederbaum:1991,
Neville_Schuurman:2020, Domcke_Woywod:1993} integration of the nonadiabatic
couplings,\cite{Baer:1975, Das_Baer:2011, Richings_Worth:2015,
Sadygov_Yarkony:1998, Esry_Sadeghour:2003} use of the molecular
properties,\cite{Mulliken:1952, Hush:1967, Cave_Newton:1997,
Werner_Meyer:1981, Yarkony:1998, Hirsch_Petrongolo:1990, Peric_Buenker:1990}
and construction of regularized quasidiabatic states.\cite{Thiel_Koppel:1999,
Koppel_Mahapatra:2001, Koppel_Schubert:2006} We have chosen the
regularized diabatization scheme because it is simple to implement and because
it removes the conical intersection singularity reliably and
efficiently.\cite{Koppel:2004} The choice of quasidiabatization affects the
magnitude of the residual couplings and, therefore, their importance for the
accuracy of nonadiabatic simulations.\cite{Choi_Vanicek:2020}

The initial state can be propagated with $\hat{\mathbf{H}}_{\text{qd-exact}}$
instead of $\hat{\mathbf{H}}_{\mathrm{ad}}$ to obtain the solution
\begin{equation}
\bm{\psi}_{\text{qd-exact}}(t) := \mathbf{S}(\hat{Q})^{\dagger} \hat
{\mathbf{U}}_{\text{qd-exact}}(t) \mathbf{S}(\hat{Q}) \bm{\psi}(0),
\label{tdse_solution_qd}%
\end{equation}
which is equivalent to $\bm{\psi}_{\mathrm{ad}}(t)$. However, it is much more
common to use the simpler approximate quasidiabatic Hamiltonian
\begin{equation}
\hat{\mathbf{H}}_{\text{qd-approx}} = \frac{\hat{P}^{2}}{2M}\mathbf{1} +
\mathbf{V}_{\mathrm{qd}}(\hat{Q}). \label{eq:mol_H_approx}%
\end{equation}
This approximation is typically justified only heuristically by referring to
the ``small'' magnitude of the residual nonadiabatic couplings. Nevertheless,
the solution
\begin{align}
\bm{\psi}_{\text{qd-approx}}(t)  &  := \mathbf{S}(\hat{Q})^{\dagger}
\hat{\mathbf{U}}_{\text{qd-approx}}(t) \mathbf{S}(\hat{Q})
\bm{\psi}(0)\nonumber\\
&  \approx\bm{\psi}_{\text{qd-exact}}(t), \label{eq:psi_approx}%
\end{align}
obtained using $\hat{\mathbf{H}}_{\text{qd-approx}}$ is not exact and would
still be only approximate even if evaluated numerically exactly.

We have numerically compared the three different solutions,
$\bm{\psi}_{\mathrm{ad}}(t)$, $\bm{\psi}_{\text{qd-exact}}(t)$, and
$\bm{\psi}_{\text{qd-approx}}(t)$ using the archetypal quadratic $E \otimes e$
Jahn--Teller model.\cite{book_Bersuker_Polinger:2012, Thiel_Koppel:1999} In
the model, two electronic states labeled by $n=1$ and $n=2$ are coupled by
doubly degenerate normal modes $Q_{1}$ and $Q_{2}$. We express the potential
energy surface in polar coordinates---the radius $\rho(Q) := \sqrt{Q_{1}^{2} +
Q_{2}^{2}}$ and polar angle $\phi(Q) := \arctan{(Q_{2}/Q_{1})}$ in the space
of the degenerate normal modes.\cite{book_Bersuker_Polinger:2012,
Thiel_Koppel:1999} Also, we work in natural units (n.u.) by setting $k = M =
\hbar= 1$ n.u., where $M$ is the mass associated with the degenerate normal
modes, and $\hbar\omega= \hbar\sqrt{k/M} = 1$ n.u. is a quantum of the
vibrational energy of these modes.\cite{book_Bersuker_Polinger:2012,
Bersuker:2001}

The adiabatic potential energy surfaces are $V_{1}(Q)=V_{+}(Q)$ and
$V_{2}(Q)=V_{-}(Q)$, where $V_{\pm}(Q):=E_{0}(Q)\pm E_{\mathrm{cpl}}(Q)$
depends on the harmonic potential energy $E_{0}(Q):=k\rho(Q)^{2}/2$ and
Jahn--Teller coupling\cite{book_Bersuker_Polinger:2012, Thiel_Koppel:1999}
\begin{equation}
E_{\mathrm{cpl}}(Q):=\rho(Q)[c_{1}^{2}+2c_{1}c_{2}\rho(Q)\cos{3\phi(Q)}%
+c_{2}^{2}\rho(Q)^{2}]^{1/2}. \label{eq:E_cpl_JT}%
\end{equation}
The nonadiabatic vector coupling
is\cite{book_Bersuker_Polinger:2012,Thiel_Koppel:1999}
\begin{equation}
\mathbf{F}_{\mathrm{ad}}(Q)=-i%
\begin{pmatrix}
\nabla\alpha(Q) & e^{2i\alpha(Q)}\nabla\theta(Q)\\
e^{-2i\alpha(Q)}\nabla\theta(Q) & -\nabla\alpha(Q)
\end{pmatrix}
\label{eq:F_ad_JT}%
\end{equation}
with $\alpha(Q):=\phi(Q)/2$ and
\begin{equation}
\theta(Q):=\frac{1}{2}\arctan{\frac{c_{1}\rho(Q)\sin{\phi(Q)}-c_{2}\rho
(Q)^{2}\sin{2\phi(Q)}}{c_{1}\rho(Q)\cos{\phi(Q)}+c_{2}\rho(Q)^{2}\cos
{2\phi(Q)}}}; \label{eq:theta_JT}%
\end{equation}
in our study, the coupling coefficients were $c_{1}=1$ n.u. and $c_{2}=0.25$
n.u. Unlike the potential energy $\mathbf{V}_{\mathrm{ad}}(Q)$,
the nonadiabatic couplings $\mathbf{F}_{\mathrm{ad}}(Q)$ and $\mathbf{G}%
_{\mathrm{ad}}(Q)$ are affected by overall phases of the adiabatic states. One
of the most standard choices\cite{book_Bersuker_Polinger:2012,
Thiel_Koppel:1999} of $A_{n}(Q)$ results in zero diagonal elements of
$\mathbf{F}_{\mathrm{ad}}(Q)$ and double-valued adiabatic states (see
Sec.~S1 of the supplementary material); instead, we have chosen the phases
$A_{n}(Q)$ so that the adiabatic states are single-valued and $\mathbf{F}%
_{\mathrm{ad}}(Q)$ contains nonzero diagonal elements [see
Eq.~(\ref{eq:F_ad_JT})]. The relationship
\begin{equation}
\mathbf{G}_{\mathrm{ad}}(Q)=\nabla\cdot\mathbf{F}_{\mathrm{ad}}(Q)+\mathbf{F}%
_{\mathrm{ad}}(Q)^{2} \label{eq:nonad_scalar_coupling}%
\end{equation}
holds exceptionally in the Jahn--Teller model and other systems in which a
finite number of states represents the system exactly in both the adiabatic
and diabatic representations. Relationship~(\ref{eq:nonad_scalar_coupling})
allows us to re-express Hamiltonian~(\ref{eq:mol_H_adiab}) in a simpler form
\begin{equation}
\hat{\mathbf{H}}_{\mathrm{ad}}=\frac{1}{2M}[\hat{P}\mathbf{1}-i\hbar
\mathbf{F}_{\mathrm{ad}}(\hat{Q})]^{2}+\mathbf{V}_{\mathrm{ad}}(\hat{Q}),
\label{eq:mol_H_adiab_simple}%
\end{equation}
which would generally only hold for $S\rightarrow\infty$.

The exact quasidiabatic Hamiltonian
\begin{align}
\hat{\mathbf{H}}_{\text{qd-exact}}  &  =\mathbf{S}(\hat{Q})\hat{\mathbf{H}%
}_{\mathrm{ad}}\mathbf{S}(\hat{Q})^{\dagger}\nonumber\\
&  =\frac{1}{2M}[\hat{P}\mathbf{1}-i\hbar\mathbf{F}_{\mathrm{qd}}(\hat
{Q})]^{2}+\mathbf{V}_{\mathrm{qd}}(\hat{Q}) \label{eq:H_qd_simple}%
\end{align}
is obtained from Hamiltonian~(\ref{eq:mol_H_adiab_simple}) using the adiabatic
to quasidiabatic transformation matrix\cite{Thiel_Koppel:1999,
Koppel_Mahapatra:2001, Koppel_Schubert:2006}
\begin{equation}
\mathbf{S}(Q)=\frac{1}{\sqrt{2}}%
\begin{pmatrix}
e^{-2i\alpha(Q)} & 1\\
1 & -e^{2i\alpha(Q)}%
\end{pmatrix}
. \label{eq:S_JT}%
\end{equation}
In Eq.~(\ref{eq:H_qd_simple}), the (no longer diagonal) quasidiabatic
potential energy matrix is
\begin{align}
\mathbf{V}_{\mathrm{qd}}(Q)  &  =\mathbf{S}(Q)\mathbf{V}_{\mathrm{ad}%
}(Q)\mathbf{S}(Q)^{\dagger}\nonumber\\
&  =%
\begin{pmatrix}
E_{0}(Q) & E_{\mathrm{cpl}}(Q)e^{-2i\alpha(Q)}\\
E_{\mathrm{cpl}}(Q)e^{2i\alpha(Q)} & E_{0}(Q)
\end{pmatrix}
, \label{eq:V_qd_JT}%
\end{align}
and the residual nonadiabatic coupling is\cite{Pacher_Koppel:1989}
\begin{align}
\mathbf{F}_{\mathrm{qd}}(Q)  &  =\mathbf{S}(Q)\mathbf{F}_{\mathrm{ad}%
}(Q)\mathbf{S}(Q)^{\dagger}+\mathbf{S}(Q)\nabla\mathbf{S}(Q)^{\dagger
}\nonumber\\
&  =-i\nabla\theta_{-}(Q)%
\begin{pmatrix}
1 & 0\\
0 & -1
\end{pmatrix}
, \label{eq:F_qd_JT}%
\end{align}
where $\theta_{\pm}(Q):=\theta(Q)\pm\alpha(Q)$ [$\theta_{+}(Q)$ will be used below].

\begin{figure}
[tbp]\includegraphics{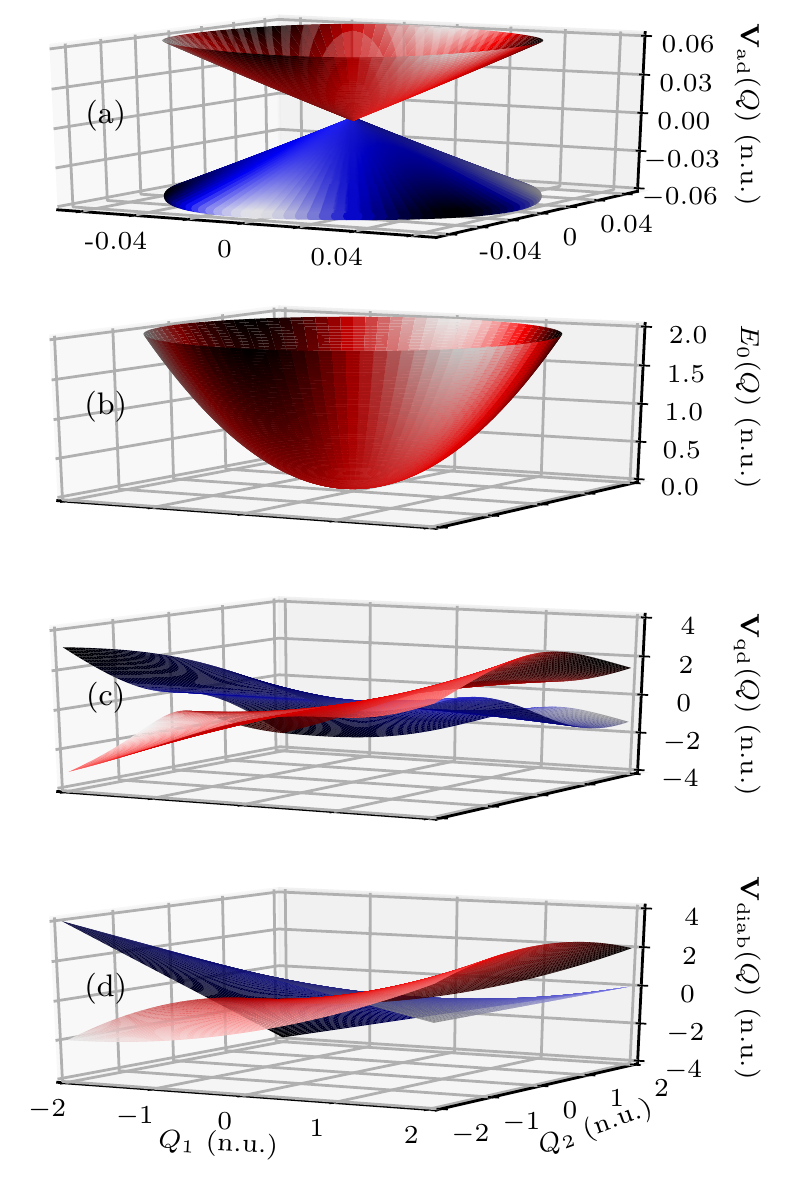}
\caption{Potential energy surfaces in the $E \otimes e$ Jahn--Teller model around the conical intersection $Q = 0$. (a) Adiabatic potential energy surfaces $V_{1}(Q) = V_{+}(Q)$ (red) and $V_{2}(Q) = V_{-}(Q)$ (blue); the two elements of the diagonal matrix $\mathbf{V}_{\mathrm{ad}}(Q)$ touch each other at the conical intersection. (b) Diagonal elements $[\mathbf{V}_{\mathrm{qd}}(Q)]_{11} =[\mathbf{V}_{\mathrm{qd}}(Q)]_{22} = [\mathbf{V}_{\mathrm{diab}}(Q)]_{11}= [\mathbf{V}_{\mathrm{diab}}(Q)]_{22}=E_{0}(Q)$ of the quasidiabatic and strictly diabatic potential energy matrices are shown together because they are all equal. (c)--(d) Off-diagonal couplings of (c) the quasidiabatic [$\mathbf{V}_{\mathrm{qd}}(Q)$] and (d) strictly diabatic [$\mathbf{V}_{\mathrm{diab}}(Q)$] potential energy matrix; the real part $\mathrm{Re}[\mathbf{V}_{i}(Q)]_{12} = \mathrm{Re}[\mathbf{V}_{i}(Q)]_{21}$ is in red, and the imaginary part $\mathrm{Im}[\mathbf{V}_{i}(Q)]_{12} = -\mathrm{Im}[\mathbf{V}_{i}(Q)]_{21}$ is in blue for $i \in \{\mathrm{qd}, \mathrm{diab} \}$. }\label{fig:pes}%

\end{figure}

In realistic systems, Eqs.~(\ref{eq:nonad_scalar_coupling}) and
(\ref{eq:mol_H_adiab_simple}) would only hold if an infinite number of
electronic states were considered. Likewise, the strictly diabatic Hamiltonian
does not, in general, exist unless $S\rightarrow\infty$%
.\cite{Mead_Truhlar:1982, Pacher_Koppel:1989} Yet, due to its exceptional
form, the Jahn--Teller Hamiltonian can be strictly
diabatized\cite{Thiel_Koppel:1999} into a separable Hamiltonian
\begin{equation}
\hat{\mathbf{H}}_{\mathrm{diab}}=\frac{\hat{P}^{2}}{2M}\mathbf{1}%
+\mathbf{V}_{\mathrm{diab}}(\hat{Q}), \label{eq:H_ref_JT}%
\end{equation}
obtained by replacing the transformation matrix $\mathbf{S}(Q)$ in
Eq.~(\ref{eq:S_JT}) with
\begin{equation}
\mathbf{T}(Q)=\frac{1}{\sqrt{2}}%
\begin{pmatrix}
e^{-i\theta_{+}(Q)} & e^{-i\theta_{-}(Q)}\\
e^{i\theta_{-}(Q)} & -e^{i\theta_{+}(Q)}%
\end{pmatrix}
; \label{eq:T_JT}%
\end{equation}
the diabatic potential energy matrix is
\begin{align}
\mathbf{V}_{\mathrm{diab}}(Q)  &  =\mathbf{T}(Q)\mathbf{V}_{\mathrm{ad}%
}(Q)\mathbf{T}(Q)^{\dagger}\\
&  =%
\begin{pmatrix}
E_{0}(Q) & E_{\mathrm{cpl}}(Q)e^{-2i\theta(Q)}\\
E_{\mathrm{cpl}}(Q)e^{2i\theta(Q)} & E_{0}(Q)
\end{pmatrix}
. \nonumber\label{eq:V_ref_JT}%
\end{align}
Potential energy surfaces $\mathbf{V}_{\mathrm{ad}}(Q)$, $\mathbf{V}%
_{\mathrm{qd}}(Q)$, and $\mathbf{V}_{\mathrm{diab}}(Q)$ in the vicinity of the
conical intersection $Q=0$ are visualized in Fig.~\ref{fig:pes}. The strictly
diabatic states are only coupled by the off-diagonal elements in
$\mathbf{V}_{\mathrm{diab}}(Q)$ because the residual nonadiabatic couplings
vanish:
\begin{equation}
\mathbf{F}_{\mathrm{diab}}(Q)=\mathbf{T}(Q)\mathbf{F}_{\mathrm{ad}%
}(Q)\mathbf{T}(Q)^{\dagger}+\mathbf{T}(Q)\nabla\mathbf{T}(Q)^{\dagger}=0.
\end{equation}
Like $\mathbf{F}_{\mathrm{ad}}(Q)$ and $\mathbf{G}_{\mathrm{ad}%
}(Q)$, the transformation matrices $\mathbf{S}(Q)$ and $\mathbf{T}(Q)$ in
Eqs.~(\ref{eq:S_JT}) and (\ref{eq:T_JT}) change according to overall phases of
the adiabatic states (see Sec.~S1 of the supplementary material).

We simulated the quantum dynamics following a transition from the ground
vibrational eigenstate of the electronic state of $A$ symmetry, $V_{A}(Q) =
-E_{\mathrm{gap}} + E_{0}(Q)$, to the doubly degenerate states of the
Jahn--Teller model by choosing the initial state as\cite{Thiel_Koppel:1999}
\begin{equation}
\tilde{\bm{\psi}}(0) = \frac{e^{- \rho(Q)^{2} /2\hbar}}{\sqrt{2 \pi\hbar}}
\begin{pmatrix}
1\\
1
\end{pmatrix}
, \label{eq:psi_0_JT}%
\end{equation}
where $\tilde{\bm{\psi}}(t) = \mathbf{T}(\hat{Q})\bm{\psi}(t)$ denotes the
wavepacket in the strictly diabatic representation. To obtain the initial
wavepacket, we assumed an impulsive excitation, i.e., the validity of the
time-dependent perturbation theory and Condon approximation during the excitation.

Among various time propagation schemes,\cite{Leforestier_Kosloff:1991,
Tal-Ezer_Kosloff:1984, Park_Light:1986, book_Hairer_Wanner:2006} we chose the
geometric integrators\cite{book_Hairer_Wanner:2006, book_Lubich:2008,
book_Leimkuhler_Reich:2004} because they preserve exactly geometric properties
of the exact evolution. Second-order split-operator
algorithms,\cite{Feit_Steiger:1982, book_Lubich:2008, book_Tannor:2007}
including the TVT algorithm, preserve the linearity, norm, inner-product,
symplecticity, stability, symmetry, and time reversibility of the exact
solution.\cite{book_Hairer_Wanner:2006, book_Lubich:2008,
book_Leimkuhler_Reich:2004} The implicit midpoint method, like the closely
related trapezoidal rule (or Crank--Nicolson)
method,\cite{Crank_Nicolson:1947,McCullough_Wyatt:1971} conserves, in
addition, the energy. Both the split-operator and implicit midpoint methods
can be symmetrically composed using various recursive or direct
schemes\cite{Suzuki:1990, Yoshida:1990, Kahan_Li:1997,
book_Hairer_Wanner:2006} to obtain integrators of arbitrary even orders of
accuracy;\cite{Choi_Vanicek:2019, Roulet_Vanicek:2019} these compositions
conserve all the geometric properties conserved by the elementary methods (see
Refs.~\onlinecite{Choi_Vanicek:2019} and \onlinecite{Roulet_Vanicek:2019}, and
Sec.~S2 of the supplementary material). The
composed\cite{book_Hairer_Wanner:2006, Suzuki:1990, Yoshida:1990} TVT
split-operator algorithm was used to propagate the wavepacket with the
separable Hamiltonians ($\hat{\mathbf{H}}_{\mathrm{diab}}$ and $\hat
{\mathbf{H}}_{\text{qd-approx}}$), whereas the composed implicit midpoint
method\cite{book_Leimkuhler_Reich:2004, Crank_Nicolson:1947,
McCullough_Wyatt:1971} was employed for propagations with the nonseparable
Hamiltonians ($\hat{\mathbf{H}}_{\mathrm{ad}}$ and $\hat{\mathbf{H}%
}_{\text{qd-exact}}$). Both integrators were composed using the
optimal\cite{Kahan_Li:1997} eighth-order scheme, which, when combined with a
small time step $\Delta t=1/(40\omega)=0.025$ n.u., led to time discretization
errors negligible to the errors due to the use of different forms of the
Hamiltonian (see Sec.~S3 of the supplementary material).

On a grid of infinite range and density, nonadiabatic quantum dynamics
simulated using $\hat{\mathbf{H}}_{\mathrm{diab}}, \hat{\mathbf{H}%
}_{\mathrm{ad}}$, and $\hat{\mathbf{H}}_{\text{qd-exact}}$ would be identical.
Therefore, the comparison of these Hamiltonians is only meaningful for a
specific finite grid; we used a uniform grid of $64 \times64$ points defined
between $Q_{l} = -10$ n.u. and $Q_{l} = 10$ n.u. for $l \in\{1,2\}$. Also, the
favorable form of the strictly diabatic Hamiltonian allowed us to obtain the
exact reference solution $\tilde{\bm{\psi}}_{\mathrm{ref}}(t) :=
\hat{\mathbf{U}}_{\mathrm{diab}}(t) \tilde{\bm{\psi}}(0)$ that is fully
converged in both space and time: to ensure the grid convergence of the
reference wavepacket, we used a grid of $128 \times128$ points defined between
$Q_{l} = -10\sqrt{2}$ n.u. and $Q_{l} = 10\sqrt{2}$ n.u. for $l \in\{1,2\}$.
In Sec.~S3 of the supplementary material, we show that both the spatial and
time discretization errors of $\tilde{\bm{\psi}}_{\mathrm{ref}}(t)$ are
negligible ($< 10^{-10}$). In contrast, even on an infinite grid, the
wavepacket $\tilde{\bm{\psi}}_{\text{qd-approx}}(t)$ would still not be exact
because the residual nonadiabatic couplings are ignored. Section~S3 of the
supplementary material shows that even on a grid of $64 \times64$ points, the
spatial discretization errors are only minor contributors to the total errors
of $\tilde{\bm{\psi}}_{\text{qd-approx}}(t)$.

\section{\label{sec:Results}Results and discussion}

\begin{figure}
[pbt]\includegraphics{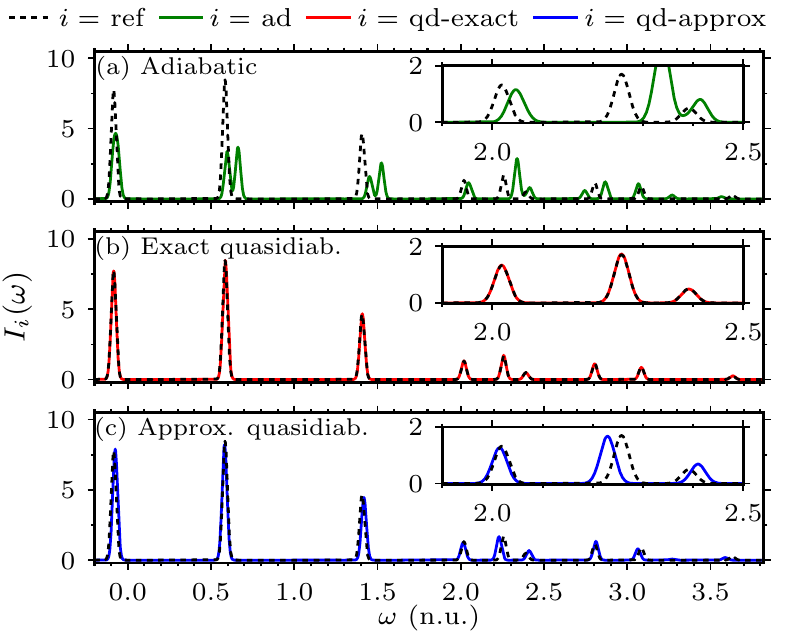}
\caption{Power spectrum computed by Fourier transforming the autocorrelation function
$\langle \tilde{\bm{\psi}}(0) | \tilde{\bm{\psi}}_{i}(t)\rangle$ of the wavepacket propagated with the (a) adiabatic ($i=\mathrm{ad}$), (b) exact quasidiabatic ($i=\textrm{qd-exact}$), or (c) approximate quasidiabatic ($i=\textrm{qd-approx}$) Hamiltonian is compared with the benchmark spectrum ($i=\mathrm{ref}$). To emulate the broadening of the peaks, the autocorrelation function was multiplied by the damping function $f(t) = \exp[-(t/t_{\mathrm{damp}})^{2}]$ with $t_{\mathrm{damp}} = 80$ n.u. before the Fourier transformation. Zoomed-in versions of the spectra are presented in the insets.}\label{fig:spectra}%

\end{figure}

\begin{figure}
[pbt]\includegraphics{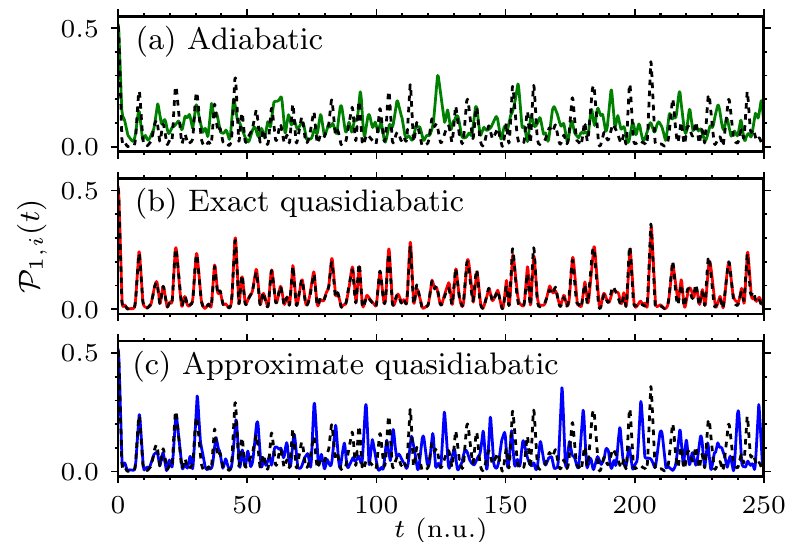}
\caption{Time dependence of the population
$\mathcal{P}_{1, i}(t) := \langle \bm{\psi}_{i}(t) | \mathbf{P}_{1} | \bm{\psi}_{i}(t) \rangle$
of the first ($n=1$) adiabatic electronic state obtained from the wavepacket propagated with the (a) adiabatic ($i=\mathrm{ad}$), (b) exact quasidiabatic ($i=\textrm{qd-exact}$), or (c) approximate quasidiabatic ($i=\textrm{qd-approx}$) Hamiltonian is compared with the benchmark population $\mathcal{P}_{1,\mathrm{ref}}(t)$; $\mathbf{P}_{n} = |n \rangle\langle n|$ is the population operator of the $n$th state. The exact norm conservation by the employed geometric integrators (see Sec.~S2 of the supplementary material) implies that the population of the second electronic state is $ \mathcal{P}_{2,i}(t) = 1 - \mathcal{P}_{1,i}(t)  $ for $i \in \{\mathrm{ref}, \mathrm{ad}, \textrm{qd-exact}, \textrm{qd-approx} \}$. Line labels are the same as in Fig.~\ref{fig:spectra}.}\label{fig:population}%

\end{figure}

\begin{figure}
[pbt]\includegraphics{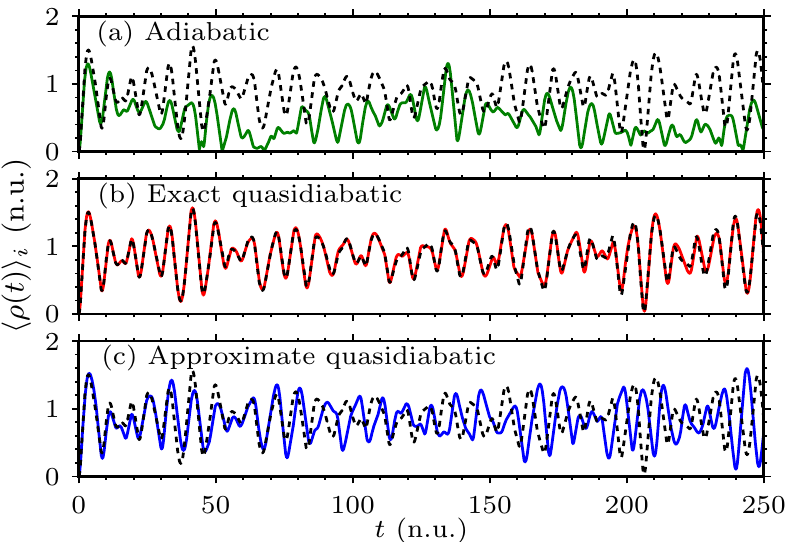}
\caption{Time dependence of the position $\langle \rho(t) \rangle_{i} := [\sum_{l=1}^{2} \langle \tilde{\bm{\psi}}_{i}(t) | \hat{Q}_{l} | \tilde{\bm{\psi}}_{i}(t) \rangle^{2}]^{1/2}$ obtained from the wavepacket propagated with the (a) adiabatic ($i=\mathrm{ad}$), (b) exact quasidiabatic ($i=\textrm{qd-exact}$), or (c) approximate quasidiabatic ($i=\textrm{qd-approx}$) Hamiltonian is compared with the benchmark position $\langle \rho(t) \rangle_{\mathrm{ref}}$. Line labels are the same as in Fig.~\ref{fig:spectra}.}\label{fig:q}%

\end{figure}

We compared the nonadiabatic quantum dynamics simulated using the adiabatic
($\hat{\mathbf{H}}_{\mathrm{ad}}$), exact quasidiabatic ($\hat{\mathbf{H}%
}_{\text{qd-exact}}$), and approximate quasidiabatic ($\hat{\mathbf{H}%
}_{\text{qd-approx}}$) Hamiltonians (see Sec.~S4 of the
supplementary material for the results obtained in the adiabatic
representation without including the geometric phase). The reference quantum
dynamics simulated using the strictly diabatic Hamiltonian $\hat{\mathbf{H}%
}_{\mathrm{diab}}$ was left out of the comparison and was only used as the
benchmark because $\hat{\mathbf{H}}_{\mathrm{diab}}$ only exists, in general,
when $S\rightarrow\infty$. Before comparing the wavepackets themselves, we
first present a comparison of three computed observables: the power spectrum
obtained by Fourier transforming the autocorrelation function $\langle
\tilde{\bm{\psi}}(0)|\tilde{\bm{\psi}}(t)\rangle$ (Fig.~\ref{fig:spectra}),
population $\mathcal{P}_{1}(t)$ of the first ($n=1$) adiabatic electronic
state (Fig.~\ref{fig:population}), and position $\langle\rho(t)\rangle$
(Fig.~\ref{fig:q}). The validity of this comparison is justified because the
time discretization errors of the presented observables and the time and
spatial discretization errors of the reference observables are negligible (see
Sec.~S3 of the supplementary material).

Panels~(a) of Figs.~\ref{fig:spectra}--\ref{fig:q} show that none of these
observables is obtained accurately with the adiabatic Hamiltonian
$\hat{\mathbf{H}}_{\mathrm{ad}}$ even if the geometric phase is
included: The positions and intensities of the peaks in the power spectrum
are inaccurate, while the population $\mathcal{P}_{1,\mathrm{ad}}(t)$ and
position $\langle\rho(t)\rangle_{\mathrm{ad}}$ deviate very rapidly from their
benchmark values. In contrast, all three presented observables are computed
extremely accurately if the wavepacket is propagated with the exact
quasidiabatic Hamiltonian [see panels~(b) of Figs.~\ref{fig:spectra}%
--\ref{fig:q}]. There is no visible difference between the observables
obtained using $\hat{\mathbf{H}}_{\text{qd-exact}}$ and the benchmark
observables obtained using $\hat{\mathbf{H}}_{\mathrm{diab}}$.

Figure~\ref{fig:spectra}(c) shows that the spectrum obtained by Fourier
transforming $\langle\tilde{\bm{\psi}}(0) | \tilde{\bm{\psi}}%
_{\text{qd-approx}}(t) \rangle$ is very similar to the benchmark spectrum; the
differences are only clearly visible in the zoomed-in version [see inset of
Fig.~\ref{fig:spectra}(c)]. For applications that do not require extremely
precise peak positions and intensities, even the spectrum obtained using
$\hat{\mathbf{H}}_{\text{qd-approx}}$ would suffice. In contrast, in
panels~(c) of Figs.~\ref{fig:population} and \ref{fig:q}, we see that both the
population and position obtained with the approximate quasidiabatic
Hamiltonian become inaccurate already after $t \approx50$ n.u. For $t > 50$
n.u., the population and position are described accurately only in simulations
using either the exact quasidiabatic or strictly diabatic Hamiltonian. Among
these Hamiltonians, however, only the exact quasidiabatic Hamiltonian exists
in general unless $S \to\infty$.\cite{Mead_Truhlar:1982, Pacher_Koppel:1989}

\begin{figure}
[ht]\includegraphics{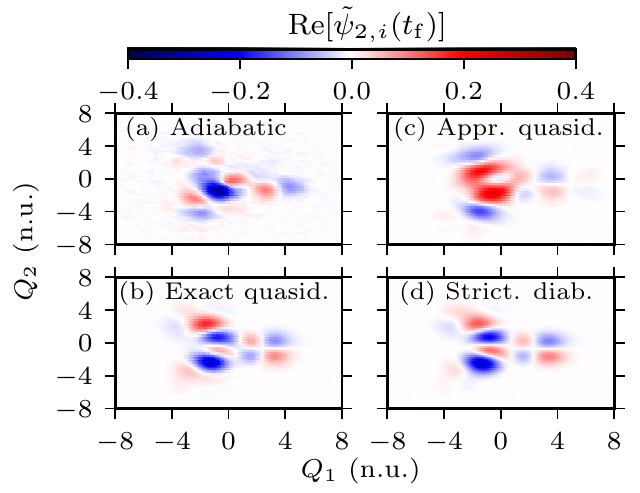}
\caption{Accuracy of the nonadiabatic quantum dynamics simulation demonstrated by comparing the
wavepackets $\tilde{\bm{\psi}}_{i}(t_{\mathrm{f}})$ at the final time $t = t_{\mathrm{f}}$ propagated
with the (a) adiabatic ($i = \mathrm{ad}$), (b) exact quasidiabatic ($i = \textrm{qd-exact}$),
(c) approximate quasidiabatic ($i = \textrm{qd-approx}$), or (d) strictly diabatic ($i = \mathrm{ref}$) Hamiltonian.
The reference wavepacket $\tilde{\bm{\psi}}_{\mathrm{ref}}(t)$ serves as the benchmark.
We only show $\mathrm{Re}[\tilde{\psi}_{2, i}(t_{\mathrm{f}})]$, i.e., the real part of the nuclear wavepacket
in the second ($n=2$) electronic state of the strictly diabatic representation.}
\label{fig:wavepacket}
\end{figure}

Some observables may be accurate, even if they are computed from a poor
wavepacket. In contrast, an accurate wavepacket ensures the accuracy of every
observable computed from it. For a more stringent comparison between the
different Hamiltonians, in Fig.~\ref{fig:wavepacket} we, therefore, display
the wavepackets $\tilde{\bm{\psi}}(t_{\mathrm{f}})$ at the final time. Whereas
$\tilde{\bm{\psi}}_{\text{qd-exact}}(t)$ resembles the exact wavepacket
$\tilde{\bm{\psi}}_{\mathrm{ref}}(t)$ closely and $\tilde{\bm{\psi}}%
_{\text{qd-approx}}(t)$ has a similar overall shape but differs in the nodal
structure and other details, $\tilde{\bm{\psi}}_{\mathrm{ad}}(t)$ is
completely different.

\begin{figure}
[ht]\includegraphics{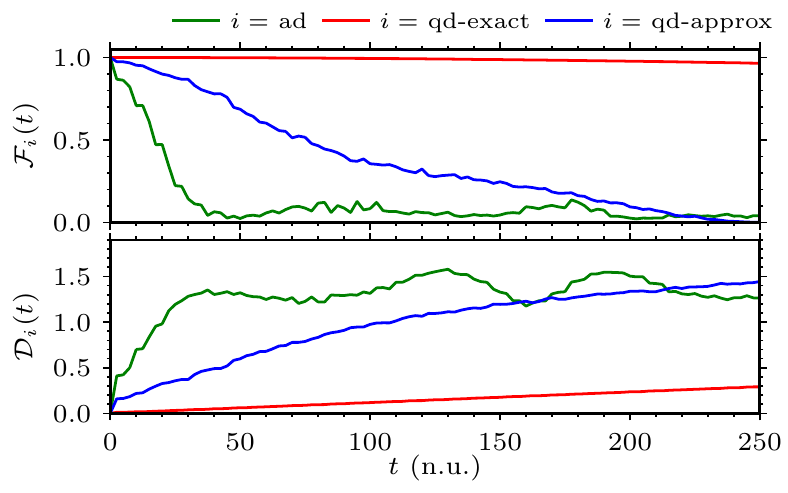}
\caption{Time dependence of the accuracy of the wavepackets propagated with different forms of the molecular Hamiltonian. The difference between $\tilde{\bm{\psi}}_{i}(t)$ and the benchmark wavepacket $\tilde{\bm{\psi}}_{\mathrm{ref}}(t)$ is measured either with the quantum fidelity\cite{Peres:1984} $\mathcal{F}_{i}(t) = |\langle \tilde{\bm{\psi}}_{\mathrm{ref}}(t)|\tilde{\bm{\psi}}_{i}(t)\rangle|^{2}$ (top panel) or distance $\mathcal{D}_{i}(t) = \| \tilde{\bm{\psi}}_{i}(t) - \tilde{\bm{\psi}}_{\mathrm{ref}}(t) \|$ (bottom panel). Fidelity and distance were computed every hundredth time step.}\label{fig:fid_diff}%

\end{figure}

For a more quantitative comparison, we measure the error of the wavepacket
$\tilde{\bm{\psi}}(t)$ using quantum fidelity\cite{Peres:1984} $\mathcal{F}%
(t):=|\langle\tilde{\bm{\psi}}_{\mathrm{ref}}(t)|\tilde{\bm{\psi}}%
(t)\rangle|^{2}$ and distance $\mathcal{D}(t):=\Vert\tilde{\bm{\psi}}%
(t)-\tilde{\bm{\psi}}_{\mathrm{ref}}(t)\Vert$ between $\tilde{\bm{\psi}}(t)$
and $\tilde{\bm{\psi}}_{\mathrm{ref}}(t)$, where $\Vert\cdot\Vert
:=\langle\cdot|\cdot\rangle^{1/2}$ denotes the norm. The quantitative
comparison, shown in Fig.~\ref{fig:fid_diff}, confirms that the quantum
dynamics simulated using the exact quasidiabatic Hamiltonian is the most
accurate: Quantum fidelity $\mathcal{F}_{\text{qd-exact}}(t)$ remains close to
its maximal value of $\mathcal{F}_{\mathrm{max}}=1$ until the final time.
Likewise, the distance $\mathcal{D}_{\text{qd-exact}}(t_{\mathrm{f}})$ at the
final time is small (although nonzero). Because $\mathcal{F}_{\text{qd-exact}%
}(t)$ stays close to its maximal value, the nonzero distance between
$\bm{\psi}_{\text{qd-exact}}(t)$ and $\bm{\psi}_{\mathrm{ref}}(t)$ is likely
to be mostly due to an overall phase difference, which does not affect
the local-in-time observables, such as population or position,
computed from the wavepackets [as shown in panels~(b) of
Figs.~\ref{fig:population} and \ref{fig:q}]. Even the approximate
quasidiabatic Hamiltonian leads to a more accurate simulation of the quantum
dynamics in the vicinity of a conical intersection than the adiabatic
Hamiltonian, which has the numerically problematic singularity of
$\mathbf{F}_{\mathrm{ad}}(Q)$ at $Q=0$.\cite{Pacher_Koppel:1989} The rapid
initial decrease of $\mathcal{F}_{\mathrm{ad}}(t)$ and increase of
$\mathcal{D}_{\mathrm{ad}}(t)$ show that the wavepacket dynamics simulated
using $\hat{\mathbf{H}}_{\mathrm{ad}}$ deviates quickly from the benchmark
solution. The decay of fidelity and increase in the distance are much more
gradual in the simulations with the approximate quasidiabatic Hamiltonian,
although both rates of change are still much faster than the rates in the
exact quasidiabatic simulation.

\section{\label{sec:conclusion}Conclusion}

We rigorously compared the suitability of different forms of the molecular
Hamiltonian for simulating the nonadiabatic quantum dynamics in the vicinity
of a conical intersection. This comparison was possible by taking advantage of
the high-order geometric integrators and exceptional existence of the strictly
diabatic Hamiltonian for the $E\otimes e$ Jahn--Teller model. The errors due
to using the different forms of the molecular Hamiltonian were measured by
comparing the fully converged exact reference simulation with simulations
performed on a slightly sparser grid using the adiabatic, exact quasidiabatic,
or approximate quasidiabatic Hamiltonian. We found that the nonadiabatic
quantum dynamics simulated using the exact quasidiabatic Hamiltonian is nearly
identical to the reference simulation obtained using the strictly diabatic
Hamiltonian. The regularized diabatization scheme\cite{Thiel_Koppel:1999,
Koppel_Mahapatra:2001, Koppel_Schubert:2006} was used for its simplicity, but
the exact quasidiabatic Hamiltonian obtained through other schemes should lead
to very similar and accurate results, as long as the quasidiabatization
removes the conical intersection singularity, because
Hamiltonian~(\ref{eq:mol_H_qd}) is exact regardless of the quasidiabatization
scheme. In contrast, the accuracy of the simulation with the approximate
quasidiabatic Hamiltonian depends on the size of the neglected residual
nonadiabatic couplings, and therefore, on the quasidiabatization
scheme.\cite{Choi_Vanicek:2020}

To return to the question posed in the title, the approximate quasidiabatic
Hamiltonian is appropriate if a quick solution of only moderate accuracy is
required because the simple separable form of this Hamiltonian allows using
more efficient time-propagation algorithms. The accuracy can be further
improved by employing more sophisticated quasidiabatization schemes, which
reduce the size of the neglected residual couplings. The adiabatic
Hamiltonian, although exact, is not suitable for simulating quantum dynamics
at a conical intersection because of the singularity of the nonadiabatic
couplings there; large errors appeared since this singularity could not be
described well on a finite grid. Although it was not subject of this study,
the adiabatic Hamiltonian is suitable for describing quantum dynamics of a
high-dimensional wavepacket moving around (i.e., not exactly through)\ a
conical intersection, especially in on-the-fly \textit{ab initio}
trajectory-based simulations, in which the adiabatic Hamiltonian is obtained
directly from electronic structure calculations, because the finite number of
trajectories propagated in such simulations are unlikely to pass directly
through conical intersection.\cite{Ben-Nun_Martinez:2000,
Curchod_Martinez:2018, Lasorne_Worth:2006, Worth_Burghardt:2004,
Saita_Shalashilin:2012} Yet, our results clearly show that the rarely used
exact quasidiabatic Hamiltonian is the most suitable form of the molecular
Hamiltonian for simulating nonadiabatic quantum dynamics directly at a conical
intersection with high accuracy. Due to the inclusion of residual nonadiabatic
couplings in the Hamiltonian, one may use any, even the simplest
quasidiabatization scheme that removes the conical intersection singularity.

\section*{Supplementary material}

See the supplementary material for the geometric phase effect in the $E\otimes
e$ Jahn--Teller model, preservation of the geometric properties by the
employed time propagation schemes, the time and spatial discretization errors
of the wavepacket and presented observables, and the nonadiabatic dynamics
simulated without including the geometric phase.

\section*{Acknowledgments}

The authors acknowledge the financial support from the European Research
Council (ERC) under the European Union's Horizon 2020 research and innovation
programme (grant agreement No. 683069 -- MOLEQULE) and thank Tomislav
Begu\v{s}i\'{c} and Nikolay Golubev for useful discussions.

\section*{Data availability}

The data that support the findings of this study are contained in the paper
and the supplementary material.

\setcounter{section}{0}
\setcounter{equation}{0}
\setcounter{figure}{0}
\setcounter{table}{0}
\renewcommand{\theequation}{S\arabic{equation}}
\renewcommand{\thefigure}{S\arabic{figure}} \renewcommand{\thesection}{S\arabic{section}}

\bigskip

\textbf{\large Supplementary material for: Which form of the molecular Hamiltonian is the
most suitable for simulating the nonadiabatic quantum dynamics at a conical intersection?}

\section{\label{appendixa}Geometric phase effect in the $E \otimes e$
Jahn--Teller model}

One way of incorporating the geometric phase effect in quantum dynamics
simulations is by representing the molecular Hamiltonian using the
double-valued adiabatic electronic states $|n^{\mathrm{dv}}(Q)\rangle$, which
change sign upon encircling a conical intersection. The sign changes of these
electronic states must be accompanied by compensating sign changes of the
associated nuclear wavefunctions $\psi_{n}^{\mathrm{dv}}(Q,t)$ in order that
the total molecular wavepacket $|\Psi(Q,t)\rangle$ be
single-valued.\cite{Mead_Truhlar:1979} However, numerically representing,
e.g., on a grid, a wavefunction that undergoes a sign change around a conical
intersection is difficult. Another way of incorporating the geometric phase
effect, according to the approach by Mead and Truhlar,\cite{Mead_Truhlar:1979}
is to multiply the double-valued adiabatic electronic states by complex phase
factors $e^{iA_{n}(Q)}$ to obtain the equivalent adiabatic states
$|n(Q)\rangle=e^{iA_{n}(Q)}|n^{\mathrm{dv}}(Q)\rangle$ that are single-valued.
Following previous work,\cite{Kendrick:2000, Juanes-Marcos_Althorpe:2005} we
use the phase $A_{n}(Q)=k_{n}\phi(Q)/2$, where $k_{n}$ must be an odd integer
to compensate for the sign change of $|n^{\mathrm{dv}}(Q)\rangle$ around a
conical intersection.

The multiplication of $|n^{\mathrm{dv}}(Q)\rangle$ by the coordinate-dependent
phase factors does not affect the diagonal adiabatic potential energy matrix
because $e^{iA_{n}(Q)}$ commutes with the electronic Hamiltonian
$\mathcal{H}_{\mathrm{e}}(Q)$:
\begin{align}
\lbrack\mathbf{V}_{\mathrm{ad}}(Q)]_{mn} &  =\langle m(Q)|\mathcal{H}%
_{\mathrm{e}}(Q)|n(Q)\rangle=\langle m^{\mathrm{dv}}(Q)|e^{-iA_{m}%
(Q)}\mathcal{H}_{\mathrm{e}}(Q)e^{iA_{n}(Q)}|n^{\mathrm{dv}}(Q)\rangle
\nonumber\\
&  =e^{-iA_{mn}(Q)}[\mathbf{V}_{\mathrm{ad}}^{\mathrm{dv}}(Q)]_{mn}%
=[\mathbf{V}_{\mathrm{ad}}^{\mathrm{dv}}(Q)]_{mn},\label{eq:V_ad_dv}%
\end{align}
where $A_{mn}(Q):=A_{m}(Q)-A_{n}(Q)$, and the last step of
Eq.~(\ref{eq:V_ad_dv}) holds since $[\mathbf{V}_{\mathrm{ad}}^{\mathrm{dv}%
}(Q)]_{mn}=\langle m^{\mathrm{dv}}(Q)|\mathcal{H}_{\mathrm{e}}%
(Q)|n^{\mathrm{dv}}(Q)\rangle$ is diagonal. In contrast, both the nuclear
wavefunctions and nonadiabatic couplings are affected by this transformation
of the adiabatic electronic states: The nuclear wavefunctions transform as
$\psi_{n}(Q,t)=e^{-iA_{n}(Q)}\psi_{n}^{\mathrm{dv}}(Q,t)$ and the nonadiabatic
couplings as%
\begin{align}
\lbrack\mathbf{F}_{\mathrm{ad}}(Q)]_{mn} &  =\langle m(Q)|\nabla
n(Q)\rangle=\langle m^{\mathrm{dv}}(Q)|e^{-iA_{m}(Q)}\nabla e^{iA_{n}%
(Q)}|n^{\mathrm{dv}}(Q)\rangle\nonumber\\
&  =e^{-iA_{mn}(Q)}[\langle m^{\mathrm{dv}}(Q)|\nabla n^{\mathrm{dv}%
}(Q)\rangle+i\nabla A_{n}(Q)\delta_{mn}]\nonumber\\
&  =e^{-iA_{mn}(Q)}\{[\mathbf{F}_{\mathrm{ad}}^{\mathrm{dv}}(Q)]_{mn}+i\nabla
A_{n}(Q)\delta_{mn}\}
\end{align}
and
\begin{align}
[&\mathbf{G}_{\mathrm{ad}}(Q)]_{mn} = \langle m(Q) | \nabla^{2} n(Q) \rangle
= \langle m^{\mathrm{dv}}(Q)|e^{-iA_{m}(Q)} \nabla^{2} e^{iA_{n}(Q)} |n^{\mathrm{dv}}(Q)\rangle \nonumber\\
& = e^{-i A_{mn}(Q)} \{ [\mathbf{G}_{\mathrm{ad}}^{\mathrm{dv}}(Q)]_{mn} + 2 i \nabla A_{n}(Q) [\mathbf{F}_{\mathrm{ad}}^{\mathrm{dv}}(Q)]_{mn}  + [i \nabla^{2} A_{n}(Q) - \nabla A_{n}(Q)^{2}]\delta_{mn} \}.
\end{align}

In the Jahn--Teller model, the adiabatic potential energy matrix is defined by
diagonalizing the strictly diabatic potential energy matrix of the
model.\cite{book_Bersuker_Polinger:2012} In
Refs.~\onlinecite{book_Bersuker_Polinger:2012, Thiel_Koppel:1999},
$\mathbf{V}_{\mathrm{diab}}(Q)$ is diagonalized by the transformation matrix
\begin{equation}
\mathbf{T}^{\mathrm{dv}}(Q)=\frac{1}{\sqrt{2}}%
\begin{pmatrix}
e^{-i\theta(Q)} & e^{-i\theta(Q)}\\
e^{i\theta(Q)} & -e^{i\theta(Q)}%
\end{pmatrix}
,\label{eq:T_dv_JT}%
\end{equation}
which is one out of a continuum of possible transformation matrices that
diagonalize $\mathbf{V}_{\mathrm{diab}}(Q)$. This standard choice
of the transformation matrix $\mathbf{T}^{\mathrm{dv}}(Q)$ yields the
adiabatic electronic states
\begin{align}
|1^{\mathrm{dv}}(Q)\rangle &  =\frac{1}{\sqrt{2}}[e^{-i\theta(Q)}|\tilde
{1}\rangle+e^{i\theta(Q)}|\tilde{2}\rangle],\nonumber\\
|2^{\mathrm{dv}}(Q)\rangle &  =\frac{1}{\sqrt{2}}[e^{-i\theta(Q)}|\tilde
{1}\rangle-e^{i\theta(Q)}|\tilde{2}\rangle]
\end{align}
that are coupled through the nonadiabatic vector couplings
\begin{equation}
\mathbf{F}_{\mathrm{ad}}^{\mathrm{dv}}(Q)=-i\nabla\theta(Q)%
\begin{pmatrix}
0 & 1\\
1 & 0
\end{pmatrix}
\end{equation}
with vanishing diagonal elements. However, the states $|n^{\mathrm{dv}%
}(Q)\rangle$ are double-valued because upon encircling a conical intersection
[i.e., as the polar angle $\phi(Q)$ changes from $-\pi$ to $\pi$], $\theta(Q)$
changes from $-\pi/2$ to $\pi/2$, causing $|n^{\mathrm{dv}}(Q)\rangle$ to
change sign. We, therefore, multiply the double-valued states $|n^{\mathrm{dv}%
}(Q)\rangle$ with phase factors $e^{iA_{n}(Q)}$, where $A_{1}(Q)=-\phi
(Q)/2=-\alpha(Q)$ and $A_{2}(Q)=\phi(Q)/2=\alpha(Q)$, to obtain the
single-valued adiabatic electronic states
\begin{align}
|1(Q)\rangle &  =\frac{1}{\sqrt{2}}[e^{-i\theta_{+}(Q)}|\tilde{1}%
\rangle+e^{i\theta_{-}(Q)}|\tilde{2}\rangle],\nonumber\\
|2(Q)\rangle &  =\frac{1}{\sqrt{2}}[e^{-i\theta_{-}(Q)}|\tilde{1}%
\rangle-e^{i\theta_{+}(Q)}|\tilde{2}\rangle].\label{eq:jt_ad_el_state}%
\end{align}
These states are single-valued because $\theta_{+}(Q)$ changes from $-\pi$ to
$\pi$ and $\theta_{-}(Q)$ starts and ends at zero as $\phi(Q)$ goes from
$-\pi$ to $\pi$. Moreover, like $\mathbf{T}^{\mathrm{dv}}(Q)$, the matrix
$\mathbf{T}(Q)$ [see Eq.~(22) of the main text], which transforms the strictly
diabatic states $|\tilde{n}\rangle$ directly into the single-valued adiabatic
states $|n(Q)\rangle$, also diagonalizes $\mathbf{V}_{\mathrm{diab}}(Q)$ [into
the same diagonal matrix as does $\mathbf{T}^{\mathrm{dv}}(Q)$] because
overall phases of the eigenvectors can be freely chosen (and here this phase
can be a function of $Q$). The two transformation matrices
$\mathbf{T}(Q)$ and $\mathbf{T}^{\mathrm{dv}}(Q)$ are related by
$\mathbf{T}(Q)=\mathbf{T}^{\mathrm{dv}}(Q)\mathbf{R}(Q)$, where $[\mathbf{R}%
(Q)]_{mn}=e^{iA_{n}(Q)}\delta_{mn}$ is the diagonal matrix that transforms the
double-valued states into the single-valued states, i.e., $|n(Q)\rangle
=\sum_{m=1}^{S}|m^{\mathrm{dv}}(Q)\rangle\lbrack\mathbf{R}(Q)]_{mn}$.
Similarly, $\mathbf{S}(Q)=\mathbf{S}^{\mathrm{dv}}(Q)\mathbf{R}(Q)$, where the
matrix
\begin{equation}
\mathbf{S}^{\mathrm{dv}}(Q)=\frac{1}{\sqrt{2}}%
\begin{pmatrix}
e^{-i\alpha(Q)} & e^{-i\alpha(Q)}\\
e^{i\alpha(Q)} & -e^{i\alpha(Q)}%
\end{pmatrix}
\label{eq:S_dv_JT}%
\end{equation}
transforms the double-valued adiabatic states into the quasidiabatic states.

\section{\label{appendixb}Conservation of geometric properties by the time
propagation schemes}

\begin{figure}
[ht]\includegraphics{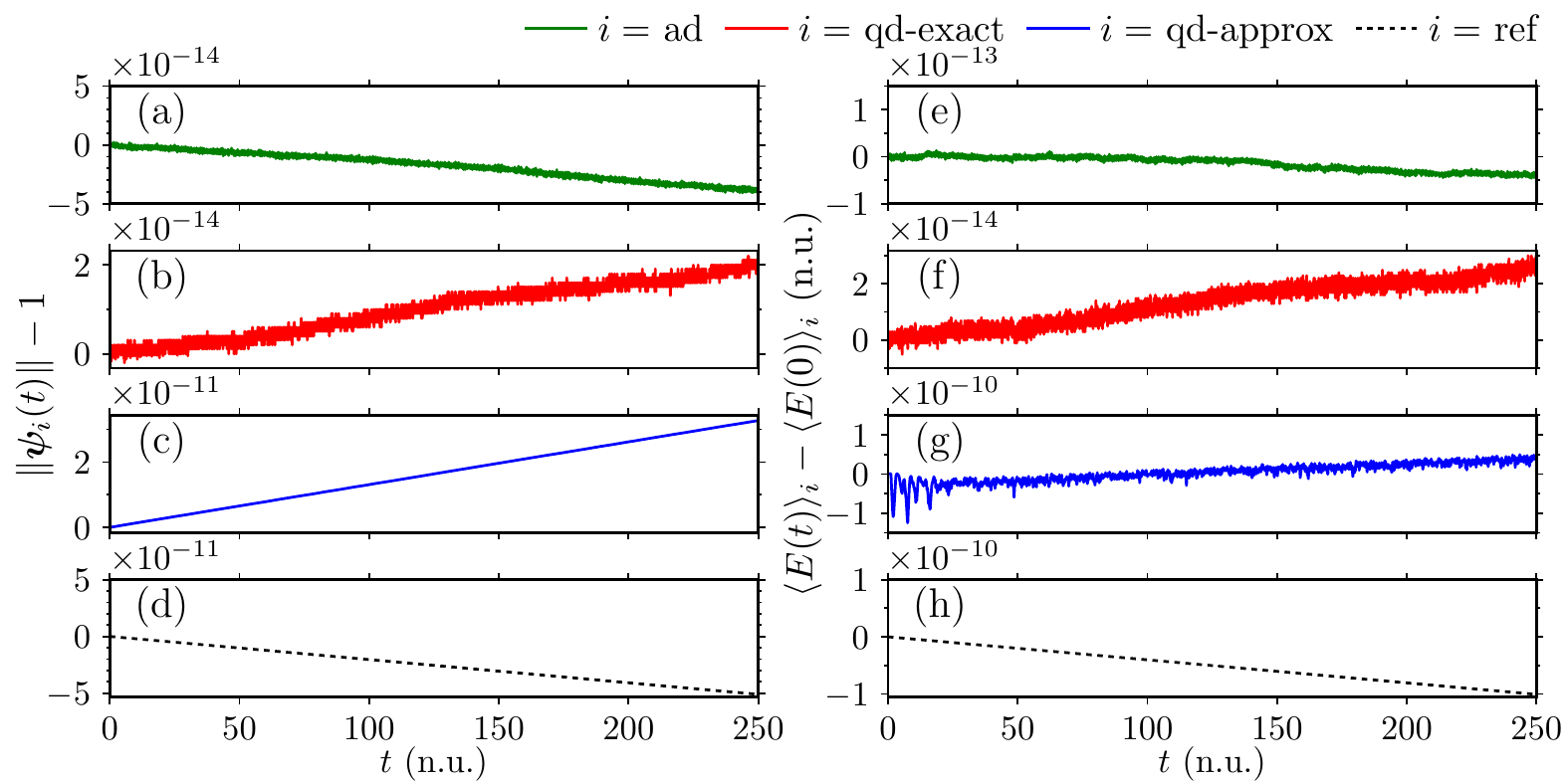}
\caption{Exact conservation of the norm $\| \bm{\psi}_{i}(t) \|$ of the wavepacket [panels (a)--(d)] and of
the expectation value $\langle E(t) \rangle_{i}$ of energy [panels (e)--(h)]. The initial value of the wavepacket
norm is $\| \bm{\psi}_{i}(0) \| = 1$, for $i \in \{ \mathrm{ad}, \textrm{qd-exact}, \textrm{qd-approx},
\mathrm{ref} \}$, and the initial values of the energy are $\langle E(0) \rangle_{\mathrm{ad}} = 2$ n.u. and
$\langle E(0) \rangle_{i} = 1$ n.u., for $i \in \{ \textrm{qd-exact}, \textrm{qd-approx}, \textrm{ref} \}$.
To evaluate the expectation value of energy, the wavepackets were transformed to the same representation
as the Hamiltonian: $\langle E(t) \rangle_{\mathrm{ad}} := \langle \bm{\psi}_{\mathrm{ad}}(t) | \hat{\mathbf{H}}_{\mathrm{ad}} | \bm{\psi}_{\mathrm{ad}}(t) \rangle$, $\langle E(t) \rangle_{i} := \langle \bm{\psi}_{i}(t) | \mathbf{S}(\hat{Q})^{\dagger} \hat{\mathbf{H}}_{i} \mathbf{S}(\hat{Q}) | \bm{\psi}_{i}(t) \rangle $ for $ i \in \{\textrm{qd-exact}, \textrm{qd-approx} \}$, and $\langle E(t) \rangle_{\mathrm{ref}} := \langle \tilde{\bm{\psi}}_{\mathrm{ref}}(t)|\hat{\mathbf{H}}_{\mathrm{diab}}|\tilde{\bm{\psi}}_{\mathrm{ref}}(t) \rangle$.}\label{fig:norm_E}%

\end{figure}

In Fig.~\ref{fig:norm_E}, we show two of the geometric properties preserved
exactly by the compositions of the implicit midpoint method: the norm of the
wavepacket and expectation value of energy. Among the two properties, only the
norm is conserved exactly by the compositions of the split-operator algorithm.
See Refs.~\onlinecite{Choi_Vanicek:2019} and \onlinecite{Roulet_Vanicek:2019}
for complete analytical and numerical demonstrations of the conservation of
linearity, inner-product, symplecticity, stability, symmetry, and time
reversibility by symmetric compositions of the split-operator or implicit
midpoint method.

We used the optimal eighth-order composition\cite{Kahan_Li:1997} of the
split-operator algorithm\cite{Feit_Steiger:1982} to propagate the wavepacket
with the separable Hamiltonians (either the approximate quasidiabatic or
strictly diabatic Hamiltonian) and the optimal eighth-order
composition\cite{Kahan_Li:1997} of the implicit midpoint
method\cite{book_Leimkuhler_Reich:2004, McCullough_Wyatt:1971} to propagate
the wavepacket with the nonseparable Hamiltonians (the adiabatic or exact
quasidiabatic Hamiltonian). Because exact norm and energy conservation are
built into the compositions of the implicit midpoint method, these geometric
properties are conserved exactly regardless of the size of the time step.
Likewise, the norm is conserved exactly by the compositions of the
split-operator algorithm. The apparent conservation of energy by the
split-operator algorithm in Fig.~\ref{fig:norm_E} only results from using a
very small time step. In fact, it was shown in
Ref.~\onlinecite{Roulet_Vanicek:2019} that the energy conservation by the
compositions of the split-operator algorithms is not exact but follows the
order of convergence of the integrator.

\section{\label{appendixc}Time and spatial discretization errors}

\begin{figure}
[ht]\includegraphics{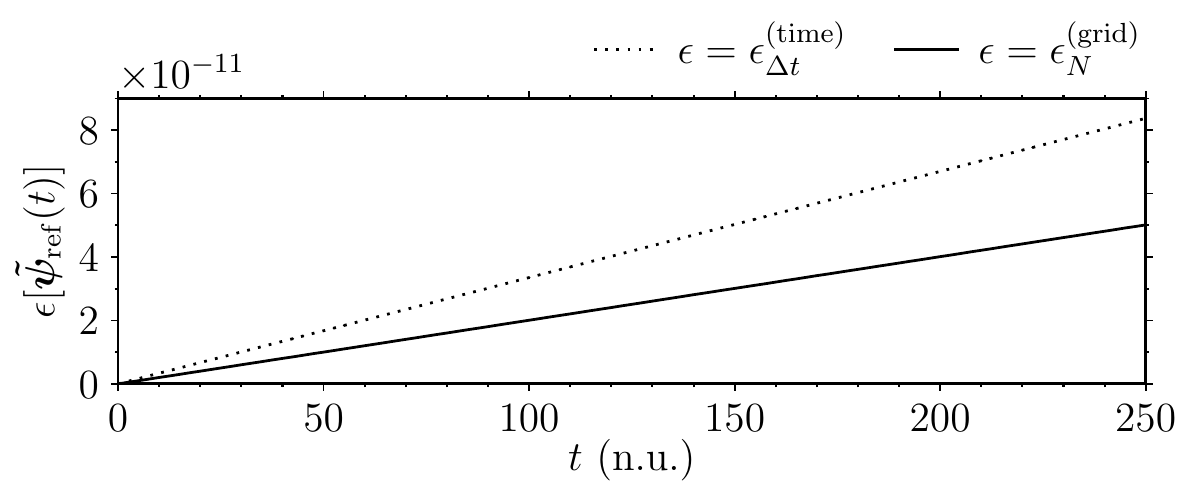}
\caption{Time discretization errors $\epsilon_{\Delta t}^{(\mathrm{time})}[\tilde{\bm{\psi}}_{\mathrm{ref}}(t)]$
(dotted line) and spatial discretization errors $\epsilon_{N}^{(\mathrm{grid})} [\tilde{\bm{\psi}}_{\mathrm{ref}}(t)]$
(solid line) of the reference wavepacket, propagated on a grid of $128 \times 128$ points ($N = 128$) using the
optimal eighth-order composition of the TVT split-operator algorithm with a time step of $\Delta t = 0.025$ n.u.
The minuscule ($< 10^{-10}$) errors indicate that  the reference
wavepacket is converged with respect to both the time step and grid size. }\label{fig:ref_error}%

\end{figure}

To quantify the errors due to different forms of the molecular Hamiltonian, an
exact reference solution is required. In principle, the quantum dynamics
simulated using any of the three exact Hamiltonians---the strictly diabatic,
exact quasidiabatic, or adiabatic Hamiltonian---can serve as a benchmark, but
the wavepacket propagated with the strictly diabatic Hamiltonian is the
easiest to converge in both time and space because of the separable form of
this Hamiltonian. We, therefore, chose the wavepacket propagated with the
strictly diabatic Hamiltonian as the benchmark. In Fig.~\ref{fig:ref_error},
we show that both the time and spatial discretization errors of propagation
with the reference Hamiltonian are negligible ($<10^{-10}$). In particular,
the numerical errors of the reference wavepacket are minuscule compared to the
errors due to different forms of the Hamiltonian (see Fig.~6 of the {main
text}).

We used the distance $\epsilon_{\Delta t}^{(\mathrm{time})}%
[\bm{\psi}(t)]:=\Vert\bm{\psi}^{(\Delta t,N)}(t)-\bm{\psi}^{(\Delta
t/2,N)}(t)\Vert$ to estimate the time discretization error of
$\bm{\psi}^{(\Delta t,N)}(t)$; here, $\bm{\psi}^{(\Delta t,N)}(t)$ denotes the
molecular wavepacket propagated to time $t$ with the time step of $\Delta t$
on a grid of $N\times N$ points. The spatial discretization errors of
$\bm{\psi}^{(\Delta t,N)}(t)$ were estimated by the distance $\epsilon
_{N}^{(\mathrm{grid})}[\bm{\psi}(t)]:=\Vert\bm{\psi}^{(\Delta t,N)}%
(t)-\bm{\psi}^{(\Delta t,2N)}(t)\Vert$. The grid of $2N\times2N$ points is
defined to be a factor of $\sqrt{2}$ wider in each dimension compared to the
grid of $N\times N$ points. Correspondingly, the grid of $2N\times2N$ points
is also a factor of $\sqrt{2}$ denser in each dimension compared to the grid
of $N\times N$ points.

\begin{figure}
[ht]\includegraphics{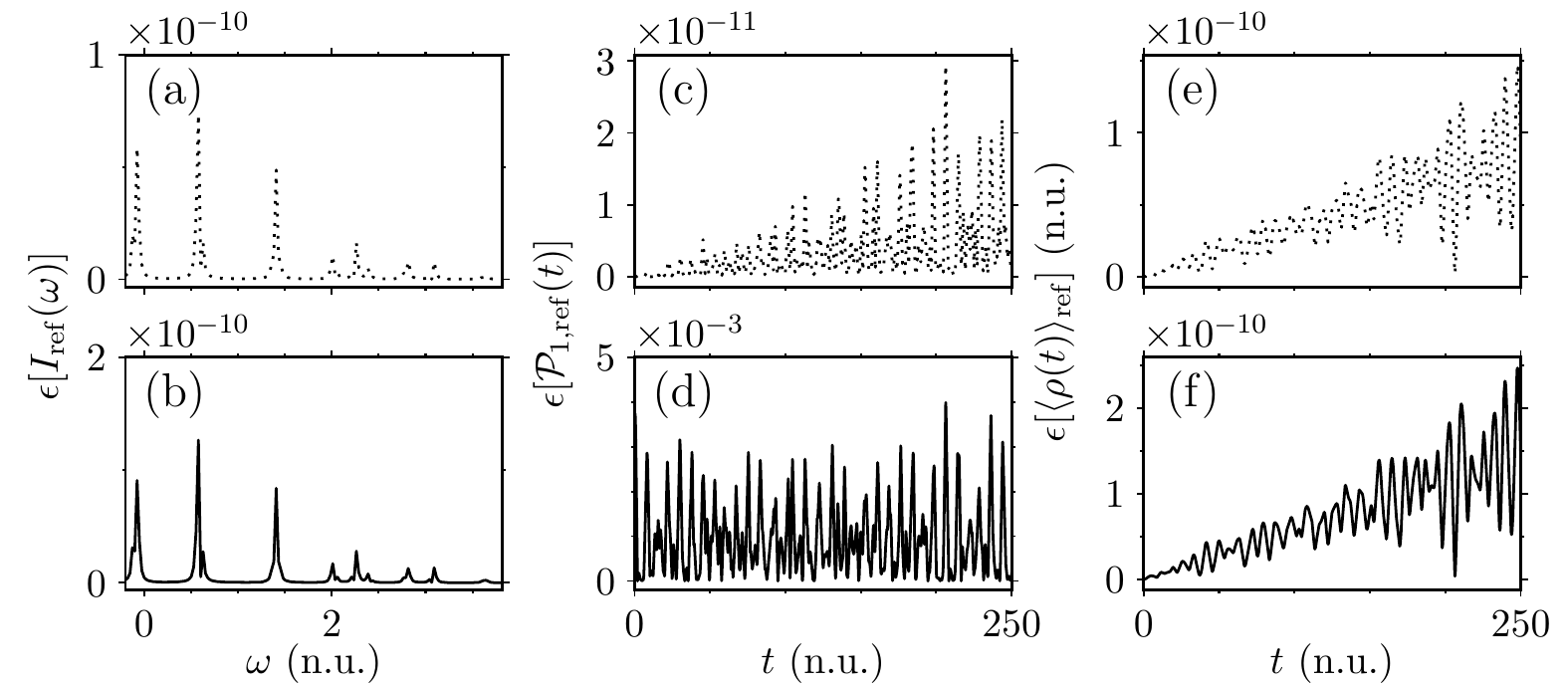}
\caption{Time discretization errors (top) and the spatial discretization errors (bottom) of the reference
observables: (a)--(b) the power spectrum, (c)--(d) population of the first ($n=1$) adiabatic electronic state,
and (e)--(f) position.
Line labels are the same as
in Fig.~\ref{fig:ref_error}.}\label{fig:ref_obs_error}
\end{figure}

For an observable $A$, we measured the time discretization errors using
$\epsilon_{\Delta t}^{(\mathrm{time})}(A):=|A^{(\Delta t,N)}-A^{(\Delta
t/2,N)}|$ and spatial discretization errors using $\epsilon_{N}%
^{(\mathrm{grid})}(A):=|A^{(\Delta t,N)}-A^{(\Delta t,2N)}|$, where
$A^{(\Delta t,N)}$ is the observable obtained from a simulation on a grid of
$N\times N$ points using the time step $\Delta t$.
Figure~\ref{fig:ref_obs_error} shows that the time and spatial discretization
errors of all reference observables are negligible ($<3\cdot10^{-10}$) except
for the spatial discretization errors of the population of the first ($n=1$)
adiabatic electronic state, which are almost entirely due to the spatial
discretization errors of the unitary transformation of the wavepacket from the
strictly diabatic to the adiabatic representation. Although larger, the
spatial discretization errors $\epsilon_{N}^{(\mathrm{grid})}[\mathcal{P}%
_{1,\mathrm{ref}}(t)]$ of the reference population are still much smaller than
the differences shown in Fig.~3 of the {main text}; even the virtually
invisible differences between $\mathcal{P}_{1,\mathrm{ref}}(t)$ and
$\mathcal{P}_{1,\text{qd-exact}}(t)$ [in Fig.~3(b) of the {main text}] are
approximately an order of magnitude greater than the spatial discretization
errors of the reference population.

\begin{figure}
[ht]\includegraphics{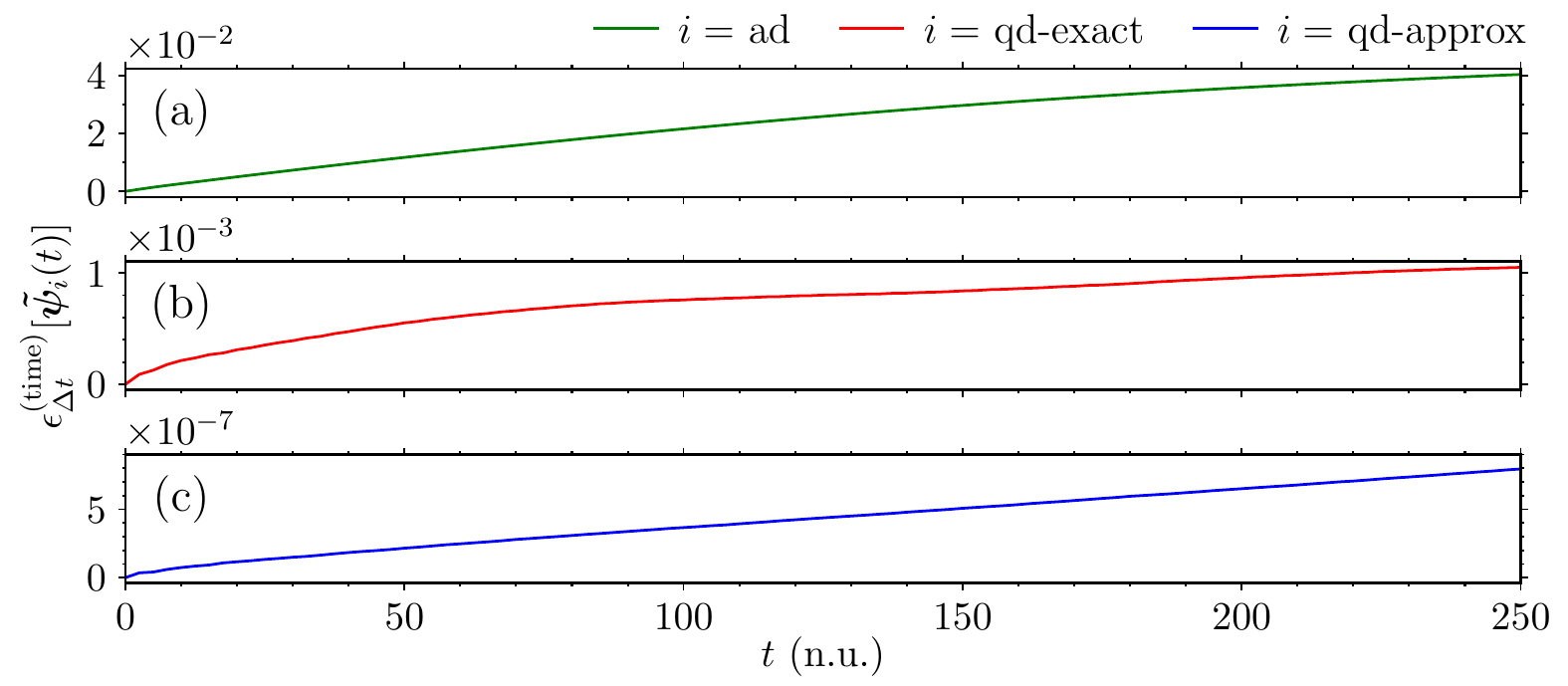}
\caption{Time discretization errors of the wavepacket propagated with
the (a) adiabatic ($i = \mathrm{ad}$), (b) exact quasidiabatic ($i = \textrm{qd-exact}$), or
(c) approximate quasidiabatic ($i = \textrm{qd-approx}$) Hamiltonian. The eighth-order composition of
the implicit midpoint method was used to propagate the wavepacket with either the adiabatic or exact
quasidiabatic Hamiltonian. The eighth-order composition of the TVT split-operator algorithm was used
to propagate the wavepacket with the approximate quasidiabatic Hamiltonian.
The time step $\Delta t = 0.025$ n.u. was used in all simulations.}\label{fig:t_error}%

\end{figure}

\begin{figure}
[pbt]\includegraphics{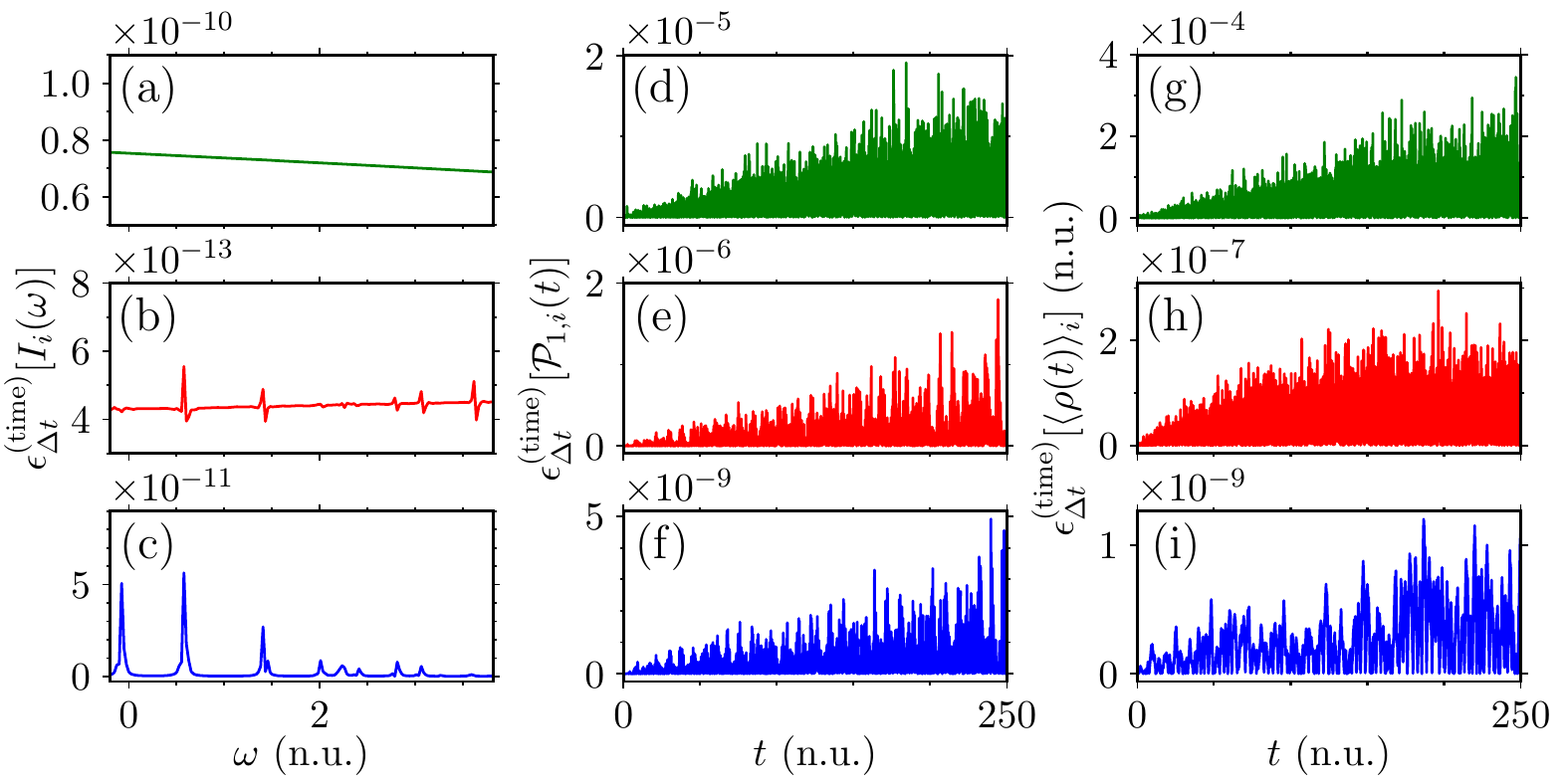}
\caption{Time discretization errors of  the power spectrum [panels (a)--(c)], population of the first ($n=1$) adiabatic
electronic state [panels (d)--(f)], and position [panels (g)--(i)] obtained in the simulation with  the adiabatic (top),
exact quasidiabatic (middle), or approximate quasidiabatic (bottom) Hamiltonian.
Line labels are the same as
in Fig.~\ref{fig:t_error}.}\label{fig:obs_t_error}
\end{figure}

Figure~\ref{fig:t_error} shows that the time discretization errors of the
propagation with the adiabatic, exact quasidiabatic, or approximate
quasidiabatic Hamiltonian are negligible (i.e., at least two orders of
magnitude smaller) compared to the errors due to different forms of the
Hamiltonian (cf. Fig.~6 of the {main text}). For the time step $\Delta
t=0.025$ n.u. employed in the simulations, the composed split-operator
algorithm, used for propagations with the separable approximate quasidiabatic
Hamiltonian, leads to much smaller time discretization errors compared to the
composed implicit midpoint method, used for propagations with the nonseparable
(adiabatic or exact quasidiabatic) Hamiltonians. In both adiabatic and
quasidiabatic simulations, the time discretization errors of the observables
(see Fig.~\ref{fig:obs_t_error}) were negligible to the errors due to the
different forms of the Hamiltonian (see Figs.~2--4 of the {main text}).

\begin{figure}
[ht]\includegraphics{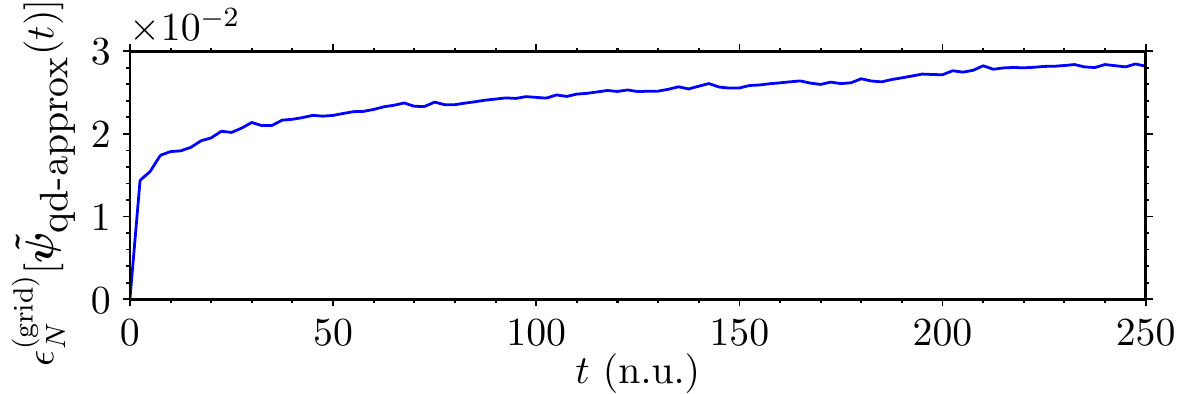}
\caption{Spatial discretization errors of the wavepacket propagated with the approximate quasidiabatic
Hamiltonian on a grid of $64 \times 64$ points ($N = 64$) using the optimal eighth-order composition of
the TVT split-operator algorithm with a time step of $\Delta t = 0.025$ n.u. The spatial discretization errors are
approximately two orders of magnitude smaller than the errors (shown in Fig.~6 of the main text) due to using the
approximate quasidiabatic Hamiltonian.}\label{fig:grid_error_approx}
\end{figure}

The spatial discretization errors of the wavepacket propagated with either the
adiabatic or exact quasidiabatic Hamiltonian are synonymous with the errors
due to the different forms of the Hamiltonian. (Recall that if the nuclear
wavefunctions were represented exactly, e.g., by using a grid of an infinite
extent and infinite density, then the wavepackets propagated with the
adiabatic or exact quasidiabatic Hamiltonian would be identical to the exact
reference wavepacket.) In contrast, the error due to using the approximate
quasidiabatic Hamiltonian consists of both the spatial discretization errors
and errors due to neglecting the residual nonadiabatic couplings. Comparing
Fig.~\ref{fig:grid_error_approx} with Fig.~6 of the {main text} shows that
neglecting the couplings is the dominant contribution.

\section{\label{appendixd}Nonadiabatic dynamics simulated without including
the geometric phase}

Hamiltonian~(16) of the main text is equivalent to the Hamiltonian
\begin{equation}
\hat{\mathbf{H}}_{\mathrm{ad}}^{\mathrm{dv}}=\mathbf{R}(\hat{Q})\hat
{\mathbf{H}}_{\mathrm{ad}}\mathbf{R}(\hat{Q})^{\dagger}=\frac{1}{2M}[\hat
{P}\mathbf{1}-i\hbar\mathbf{F}_{\mathrm{ad}}^{\mathrm{dv}}(\hat{Q}%
)]^{2}+\mathbf{V}_{\mathrm{ad}}(\hat{Q})\nonumber
\end{equation}
represented using the double-valued adiabatic states $|n^{\mathrm{dv}%
}(Q)\rangle$. However, like the electronic states themselves, the initial
state $\bm{\psi}^{\mathrm{dv}}(0)=\mathbf{R}(\hat{Q})\bm{\psi}(0)$ in this
representation is also double-valued. We examine the effect of neglecting the
geometric phase by simulating the dynamics with one of the two branches of the
double-valued state $\bm{\psi}^{\mathrm{dv}}(0)$ as the initial wavepacket.

\begin{figure}
[ht]\includegraphics{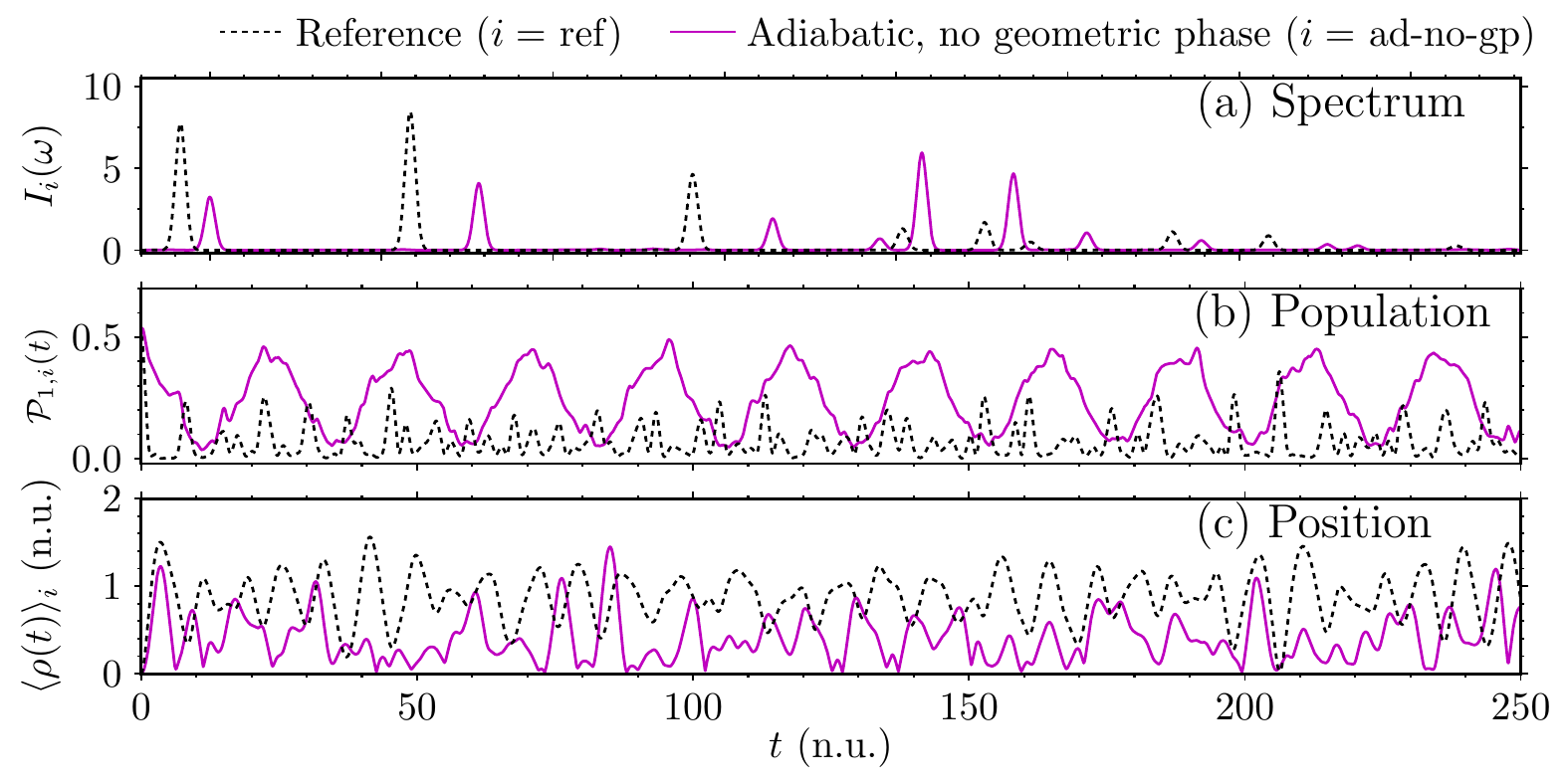}
\caption{(a) Power spectrum, (b) population, and (c) position obtained without including the geometric phase
in the adiabatic representation ($i=\textrm{ad-no-gp}$) are compared with the reference observables obtained with
the strictly diabatic Hamiltonian. Panels~(a), (b), and (c) are analogous to Figs.~2(a), 3(a), and 4(a) of the main
text, respectively.}\label{fig:obs_ad_no_GP}
\end{figure}

Figure~\ref{fig:obs_ad_no_GP} shows that the power spectrum $I_{i}(\omega)$
[panel~(a)], population $\mathcal{P}_{1,i}(\omega)$ [panel~(b)], and position
$\langle\rho(t)\rangle_{i}$ [panel~(c)] obtained in the adiabatic
representation without the geometric phase ($i=\text{ad-no-gp}$) are much less
accurate than those obtained in the adiabatic representation in which the
geometric phase is included [$i=\mathrm{ad}$, Figs.~2(a), 3(a), and 4(a) of
the main text].

\begin{figure}
[ht]\includegraphics{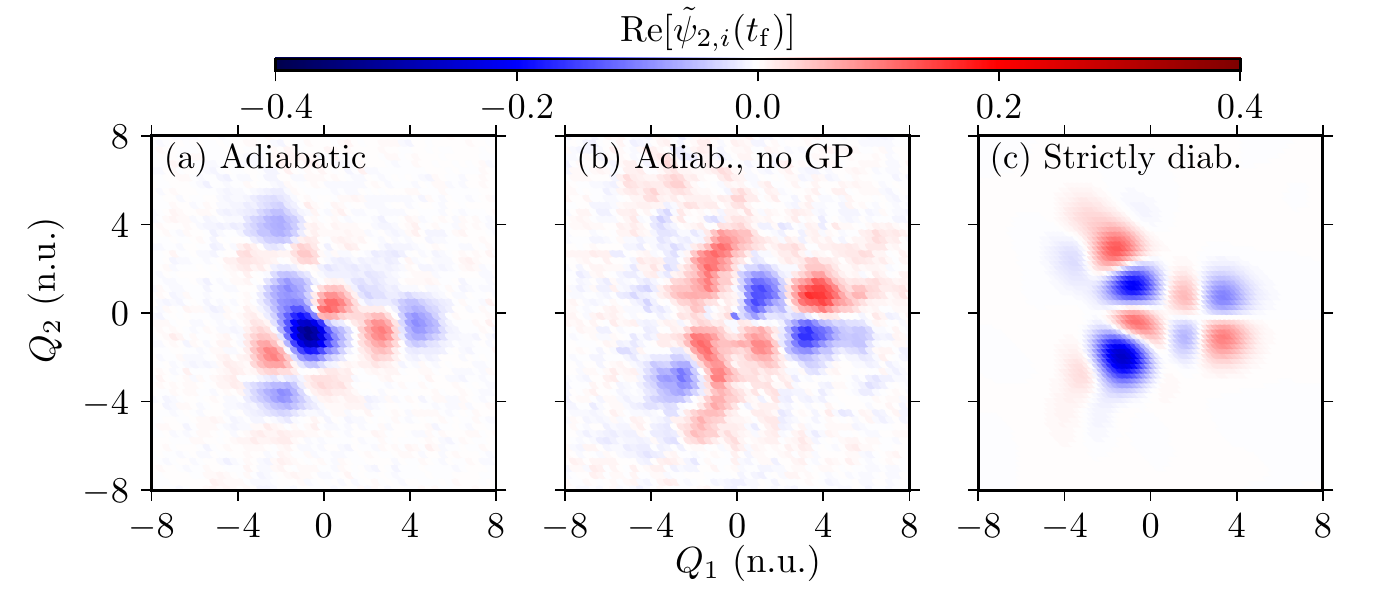}
\caption{Effect of neglecting the geometric phase demonstrated by comparing the wavepackets
$\tilde{\bm{\psi}}_{i}(t_{\mathrm{f}})$ at the final time, propagated in the adiabatic representation
with [$i=\mathrm{ad}$, panel~(a)] or without [$i=\textrm{ad-no-gp}$, panel~(b)] the geometric phase.
The benchmark wavepacket $\tilde{\bm{\psi}}_{\mathrm{ref}}(t_{\mathrm{f}})$ propagated in the strictly diabatic
representation is shown as a reference in panel~(c).}
\label{fig:wavepacket_adiab}
\end{figure}

\begin{figure}
[ht]\includegraphics{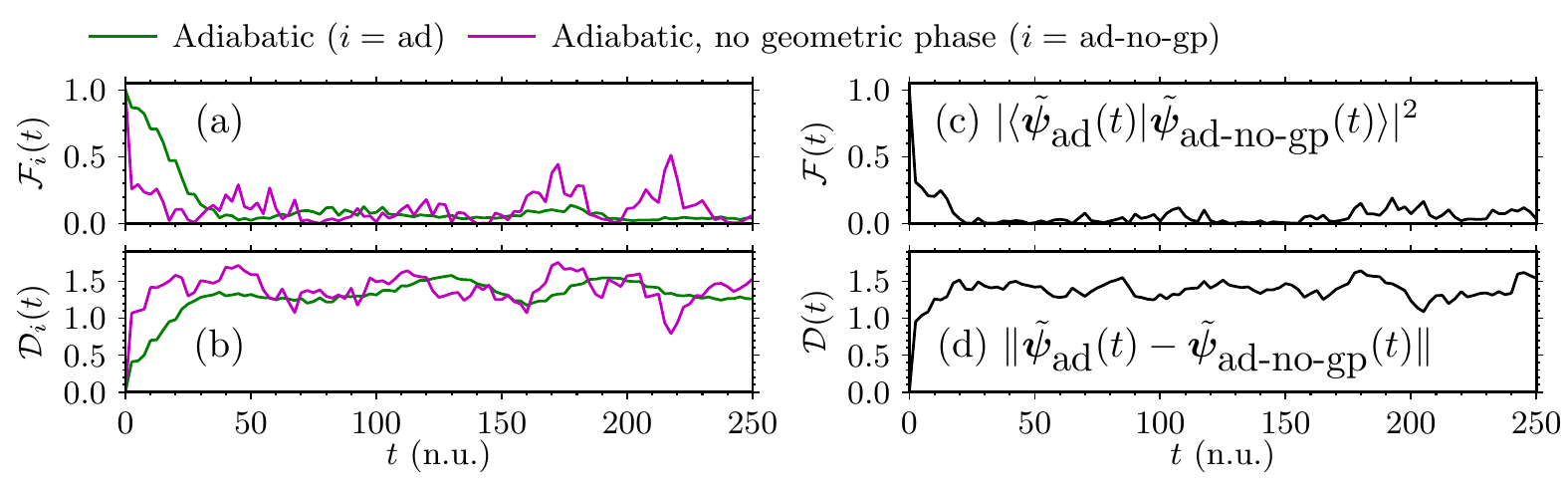}
\caption{Time dependence of the accuracy of the nonadiabatic dynamics simulated in the adiabatic
representation either with ($i=\mathrm{ad}$) or without ($i=\textrm{ad-no-gp}$) the geometric phase is
measured by (a) the quantum fidelity $\mathcal{F}_{i}(t)$ and (b) distance $\mathcal{D}_{i}(t)$
between $\tilde{\bm{\psi}}_{i}(t)$ and the benchmark wavepacket $\tilde{\bm{\psi}}_{\mathrm{ref}}(t)$.
The importance of the geometric phase is measured more directly by (c) the quantum fidelity
$\mathcal{F}(t)=| \langle \tilde{\bm{\psi}}_{\mathrm{ad}}(t) |\tilde{\bm{\psi}}_{\textrm{ad-no-gp}}(t) \rangle |^{2}$
and (d) distance $\mathcal{D}(t) = \|\tilde{\bm{\psi}}_{\mathrm{ad}}(t) - \tilde{\bm{\psi}}_{\textrm{ad-no-gp}}(t) \|$
between the wavepackets propagated in the adiabatic representation either with ($i=\mathrm{ad}$) or without
($i=\textrm{ad-no-gp}$) the geometric phase.}\label{fig:fid_diff_adiab}
\end{figure}

At the final time $t=t_{\mathrm{f}}=250$ n.u., the wavepacket propagated in
the adiabatic representation does not resemble the reference wavepacket
$\tilde{\bm{\psi}}_{\mathrm{ref}}(t_{\mathrm{f}})$ regardless of whether the
geometric phase is included or not [compare panels~(a), (b), and (c) of
Fig.~\ref{fig:wavepacket_adiab}]. We, therefore, present the time dependence
of various measures of the accuracy of the wavepacket $\tilde{\bm{\psi}}%
_{i}(t)$ propagated either with ($i=\mathrm{ad}$) or without
($i=\text{ad-no-gp}$) the geometric phase in Fig.~\ref{fig:fid_diff_adiab}.
The rapid decrease of quantum fidelity$\mathcal{F}_{\text{ad-no-gp}}(t)$
[panel~(a)] and rapid increase of distance $\mathcal{D}_{\text{ad-no-gp}}(t)$
[panel~(b)] from the reference wavepacket show that without the geometric
phase, the simulation becomes inaccurate almost immediately. The similarly
sharp changes in quantum fidelity $\mathcal{F}(t)$ and distance $\mathcal{D}%
(t)$ between the two wavepackets $\tilde{\bm{\psi}}_{i}(t)$ propagated either
with ($i=\mathrm{ad}$) or without ($i=\text{ad-no-gp}$) the geometric phase
(shown in the right-hand panels of Fig.~\ref{fig:fid_diff_adiab}) confirms
that ignoring the geometric phase is responsible for the rapid decrease in the
accuracy of $\tilde{\bm{\psi}}_{\text{ad-no-gp}}(t)$.

\clearpage

\bibliographystyle{aipnum4-2}
\bibliography{Hamiltonian_representation_CI}

\end{document}